\documentclass[lettersize,onecolumn]{IEEEtran}

\usepackage[utf8]{inputenc}
\usepackage{amsmath,amsfonts,amssymb,amsthm}
\usepackage{mathrsfs}
\usepackage[all]{xy}
\usepackage{marginnote}
\usepackage{todonotes}
\usepackage{graphicx}
\usepackage{colortbl}
\usepackage{epstopdf}
\usepackage{array}
\usepackage{nicematrix}
\usepackage{tikz}
\usepackage{verbatim}
\usepackage{arydshln}
\usepackage{cite}
\usepackage{url}
\usepackage[colorlinks=true,citecolor=blue,linkcolor=blue,urlcolor=blue]{hyperref}

\graphicspath{{./}{sections/}}
\newcommand{\ket}[1]{|#1\rangle}
\newcommand{\bra}[1]{\langle#1|}

\newcommand*{\Tr}{\operatorname{Tr}}
\newcommand*{\supp}{\operatorname{supp}}
\newcommand*{\wt}{\operatorname{wt}}
\newcommand*{\rank}{\operatorname{rank}}
\newcommand{\ketbra}[2]{|#1\rangle\!\langle#2|}

\newtheorem{theorem}{Theorem}
\newtheorem{lemma}{Lemma}
\newtheorem{definition}{Definition}
\newtheorem{corollary}{Corollary}
\theoremstyle{definition}
\newtheorem{remark}{Remark}

\newtheorem{example}{Example}

\newtheorem{proposition}{Proposition}

\DeclareFontFamily{U}{mathx}{}
\DeclareFontShape{U}{mathx}{m}{n}{<-> mathx10}{}
\DeclareSymbolFont{mathx}{U}{mathx}{m}{n}
\DeclareMathAccent{\widehat}{0}{mathx}{"70}
\DeclareMathAccent{\widecheck}{0}{mathx}{"71}

\begin{document}

\title{Classification of Small-size Quantum Secret Sharing Schemes using Uniform States}

\author{Xuhong Liu and Shuai Shao%
\thanks{Xuhong Liu and Shuai Shao are with the School of Computer Science and Technology \& Hefei National Laboratory, University of Science and Technology of China (e-mail: xuhongliu@mail.ustc.edu.cn; shao10@ustc.edu.cn).}}

\hypersetup{
    pdftitle={Classification of Small-size Quantum Secret Sharing Schemes using Uniform States},
    pdfauthor={Xuhong Liu and Shuai Shao}
}

\markboth{IEEE Transactions on Information Theory}{Liu and Shao: Classification of Small-size Quantum Secret Sharing Schemes using Uniform States}

\maketitle

	\begin{abstract}
		We study the connection between quantum secret sharing (QSS) schemes and $k$-uniform states of qubits beyond the equivalence between threshold QSS schemes and AME states. 
        Specifically, we show that $3$-uniformity is a necessary but not sufficient condition for constructing a $3$-homogeneous QSS scheme using states of qubits. 
       To the best of our knowledge, this is the first result connecting \emph{non-threshold} QSS schemes with $k$-uniform states. 
        As an application of our result, we classify QSS schemes for up to 7 players and provide explicit characterizations of their existence. 
        Our results offer new insights into the role of $k$-uniform states in the design of QSS schemes (not necessarily threshold) and provide a foundation for future classifications of QSS schemes with more complex structures.

		\end{abstract}


\section{Introduction}
	A \emph{quantum secret sharing} (QSS) scheme is a quantum cryptographic scheme for sharing a secret among $n$ players such that 
    specific subsets of players, called \emph{authorized sets}, can collaborate to reconstruct the secret using their shares. 
    In contrast, all other subsets, known as \emph{unauthorized sets}, gain no information about the secret. 
   Typically, $n \geq 3$ is assumed. Otherwise, the scheme is trivial.
    The collection of authorized sets is called the \emph{access structure} of the scheme.
    The concept of QSS schemes was first introduced by Hillery et al. for classical secrets \cite{Hillery_1999}, and later extended to quantum secrets by Cleve et al.~\cite{PhysRevLett.83.648}, who also established a connection between quantum secret sharing and quantum error-correcting codes.  
    Since then, QSS has attracted substantial research interest both theoretically \cite{PhysRevLett.117.030501,10.1007/978-981-96-0947-5_5,PhysRevA.83.042303,PhysRevA.81.052333,9674910,cryptoeprint:2014/322,PhysRevA.86.052335,PhysRevA.63.042301,PhysRevA.69.052307,bell2014experimental,PhysRevA.78.062307,HAO20113639,PhysRevLett.95.200502} and experimentally \cite{PhysRevA.63.042301,PhysRevA.69.052307,bell2014experimental,PhysRevA.78.062307,HAO20113639,PhysRevLett.95.200502,PhysRevLett.117.030501}.

    Theoretically, 
    Gottesman \cite{PhysRevA.61.042311} and Smith \cite{smith2000quantumsecretsharinggeneral} independently proved that QSS schemes can be constructed for any access structure satisfying key quantum laws, such as no-cloning and monotonicity. 
    However, their constructions often require significantly more entanglement resources than practical implementations can afford. 
    To share a single-qudit secret among $n$ players, their constructions may require preparing entangled quantum states of $m$ qudits, where $m \gg n$.
    This results in inefficient use of valuable entanglement resources, often making such QSS schemes impractical.
Thus, there is a need to design more efficient  QSS schemes. 
Specifically, the goal is to share a quantum secret using a quantum system with the smallest possible number of qudits. The ideal minimum number of qudits is clearly $n$, where each player receives a single qudit as their share. 
Unless otherwise specified, in this paper we consider QSS schemes that share a single qudit secret with $n$ players using a quantum system of $n$ qudits. 
We are concerned with determining the access structures for which such QSS schemes can exist.

A particularly important class of access structures is the \emph{$k$-threshold structure}, in which any group of at least $k$ players can reconstruct the secret, while any group of fewer than $k$ players obtains no information about it.
Equivalently, a subset is authorized if and only if its size is at least $k$.
Threshold QSS schemes are closely related to \emph{absolutely maximally entangled} (AME) states, which are special cases of \emph{$k$-uniform states}.
A \emph{$k$-uniform state} is a pure quantum state of $n$ qudits such that every reduction to $k$ qudits is maximally mixed, i.e., for any $A \subseteq [n]$ with $|A|=k$, the reduced density matrix satisfies $\rho_A=\frac{\mathbb{I}_A}{d^{|A|}}$.
An $n$-qudit AME state is a $k$-uniform state, where $k$ takes the maximal possible value $\left\lfloor \frac{n}{2} \right\rfloor$. 

The concept of $k$-uniform states was first introduced in \cite{PhysRevA.69.052330}.
A variety of combinatorial and algebraic techniques have been employed to construct such states, including Latin squares, symmetric matrices, graph states, quantum error-correcting codes, and classical error-correcting codes~\cite{PhysRevA.90.022316,PhysRevA.69.052330,7918542,PhysRevA.92.032316,PhysRevA.97.062326,helwig2013absolutelymaximallyentangledqudit,PhysRevResearch.2.033411}.
Despite  multiple construction techniques, the general question of the existence of $k$-uniform states for arbitrary parameters remains open.
$k$-uniform states have found applications in both theory and practice, including quantum error correction~\cite{PhysRevA.69.052330}, information masking~\cite{PhysRevA.104.032601}, quantum teleportation~\cite{PhysRevA.63.042301}, and holographic quantum codes~\cite{Pastawski_2015}.
In particular, AME states (the extremal case of $k$-uniform states) can be directly used to construct QSS schemes.
Indeed, there exists a one-to-one correspondence between an $n$-threshold QSS scheme for $2n-1$ players and an AME state of $2n$ qudits~\cite{PhysRevA.86.052335}.
However, AME states are relatively rare.
For qubits, AME states exist only when the number of qubits $n$ is $3, 5$ or $6$, with a unique AME state for each $n$~\cite{Huber_2018}. 

In addition to AME states, quantum error correction codes have been employed in constructing QSS schemes. 
For example, Calderbank–Shor–Steane (CSS) codes~\cite{PhysRevA.54.1098}, such as the Steane code introduced in \cite{steane1996multiple}, have been used to construct QSS schemes. The Steane code encodes one qubit into seven qubits, and can correct arbitrary single-qubit errors. Sarvepalli showed that the Steane code can be used to construct a non-threshold QSS scheme with seven players \cite{PhysRevA.81.052333}.
Moreover, the access structure of this QSS scheme is \emph{$3$-homogeneous}, i.e., every minimal authorized set in its access structure (no proper subset of which is authorized) contains exactly three players. 
Clearly, $k$-threshold QSS schemes are a special case of $k$-homogeneous QSS schemes.

The Steane code induces an eight-qubit stabilizer state $\ket{f_8} = \frac{1}{\sqrt{2}}(\ket{0}\otimes\Lambda_s(\ket{0})+\ket{1}\otimes\Lambda_s(\ket{1}))$.
The state $\ket{f_8}$ exhibits remarkable properties and has been investigated from multiple perspectives, including quantum coding~\cite{steane1996multiple}, orthogonal arrays~\cite{PhysRevA.99.042332}, and classical complexity classification~\cite{9317917}.
It is known that $\ket{f_8}$ is a $3$-uniform state, and it has also been shown to attain the upper bound of the \emph{Quantum Extremal Number} (Qex number)~\cite{PhysRevA.111.052410}.
Thus, through the Steane code, a $3$-homogeneous (non-threshold) QSS scheme can be connected with a $3$-uniform state.
This observation naturally raises the following question:
\emph{Can $k$-uniform states be used to characterize $k$-homogeneous QSS schemes (not just $k$-threshold schemes), with AME states of $2n$ qudits specifically characterizing threshold QSS schemes with $2n-1$ players?}

\subsection*{Our contributions}
In this paper, we take a first step toward answering the above question. We focus on quantum systems of qubits and explore the connection between QSS schemes and $k$-uniform states beyond the known equivalence between threshold QSS schemes and AME states.
We first show that $3$-homogeneous QSS schemes imply $3$-uniform states. To the best of our knowledge, this is the first result connecting \emph{non-threshold $k$-homogeneous} QSS schemes with $k$-uniform states. Although this result is established only for $3$-homogeneous QSS schemes, the restriction to $3$ is close to optimal. In particular, we construct a $5$-homogeneous QSS scheme whose corresponding state is only $3$-uniform rather than $5$-uniform.
The converse of our result does not hold either. We show that the $3$-uniform state given in~\cite{PhysRevA.111.052410}, which achieves the maximum Qex number, cannot be used to construct a $3$-homogeneous QSS scheme. Therefore, $3$-uniformity is a necessary but not sufficient condition for a quantum state of qubits to give rise to a $3$-homogeneous QSS scheme.

This necessary condition also provides a useful tool for classifying QSS schemes with a small number of players. For QSS schemes with at most $7$ players, we show that, up to renaming the players, there are only two valid schemes: the $3$-threshold QSS scheme with $5$ players corresponding to the unique AME state of $6$ qubits, and the $3$-homogeneous QSS scheme with $7$ players corresponding to the $3$-uniform state $\ket{f_8}$. Our classification results are summarized in the table below.

	\begin{table}[ht]
    \centering
    \setlength{\tabcolsep}{32pt}
    \renewcommand{\arraystretch}{1.2}
    \begin{tabular}{|c|c|}
        \hline
        Number of players & Existence\\
        \hline
        $n=3$ & no scheme exists \\
        \hline
        $n=4$ & no scheme exists \\
        \hline
        $n=5$ & a unique $3$-threshold scheme corresponding to the unique AME state of $6$ qubits\\
        \hline
        $n=6$ & no scheme exists \\
        \hline
        $n=7$  & a unique $3$-homogeneous scheme corresponding to the $3$-uniform state $\ket{f_8}$ \\
        \hline
    \end{tabular}
    \vspace{1.6ex}
    \caption{Characterization of quantum secret sharing schemes with small numbers of players.}\label{tab}
\end{table}
\vspace{-1ex}

Compared with the previous work establishing the equivalence between AME states and threshold QSS schemes~\cite{helwig2013absolutelymaximallyentangledstates}, our work uses several additional ideas and techniques:
 While the proof in~\cite{helwig2013absolutelymaximallyentangledstates} relies mainly on entropy inequalities, our approach also makes essential use of the combinatorial structure of access structures. In particular, we establish several structural properties of these access structures and use combinatorial tools such as the Fano plane. In addition,  building on the connection between $3$-uniform states and $3$-homogeneous QSS schemes, we reduce the existence of certain QSS schemes to the existence of specific $3$-uniform states. We then use shadow enumerators and related inequalities~\cite{817508} to prove the non-existence of several classes of such states.

Our results provide a new perspective on the role of $k$-uniform states in the study of QSS schemes. The classification of QSS schemes with small numbers of players provides a foundation for studying QSS schemes with more complicated access structures and may also help guide the design of QSS schemes in experimental settings.

The paper is organized as follows. Section~\ref{sec:pre} introduces preliminaries.  In
Section~\ref{sec:uniform-qss-comparison}, we study the connections between $k$-homogeneous QSS schemes and $k$-uniform states. In Section~\ref{4.1}, we present the classification of QSS schemes with at most seven players. 
Section~\ref{c} concludes with further directions.

	\section{Preliminaries}\label{sec:pre}
	\subsection{Notation and definitions}
We use standard Dirac notation. For a finite-dimensional Hilbert space
\(\mathcal H\), let \(\mathcal B(\mathcal H)\) denote the space of linear
operators on \(\mathcal H\), and let \(D(\mathcal H)\) denote the set of density
operators on \(\mathcal H\). A pure state \(\ket{\psi}\) is identified with the
rank-one density operator \(\ket{\psi}\bra{\psi}\). For a multipartite state
\(\rho\), we write \(\rho_A\) for the reduced state on subsystem \(A\).

The von Neumann entropy of a density operator \(\rho\) is
\(S(\rho):=-\Tr(\rho\log\rho)\). When the underlying state is
clear, we write \(S(A)\) for \(S(\rho_A)\). If
\(\rho\in D(\mathbb C^d)\), then \(S(\rho)\leq \log d\), with equality if and
only if \(\rho=\mathbb{I}_d/d\).

We shall use the following standard entropy inequalities. For disjoint
subsystems \(A\) and \(B\),
\begin{align}
    S(A\cup B) &\leq S(A)+S(B), \label{tri}\\
    S(A\cup B) &\geq |S(A)-S(B)|. \label{AL}
\end{align}
The first inequality is subadditivity, and the second is the Araki--Lieb
inequality. We also use strong subadditivity: for arbitrary subsystems \(A\)
and \(B\),
\begin{equation}
    S(A)+S(B) \geq S(A\cup B)+S(A\cap B).
    \label{strong}
\end{equation}
These inequalities are standard; see, e.g.,~\cite{Nielsen_Chuang_2010}.

\begin{definition}[Mutual information]\label{mutual}
For two subsystems \(A\) and \(B\), the mutual information is
\(I(A:B):=S(A)+S(B)-S(A\cup B)\).
\end{definition}

For a positive integer \(n\), let \([n]:=\{1,\ldots,n\}\). We write
\(2^{[n]}\) for the power set of \([n]\), \(\binom{[n]}{k}\) for the family of
subsets \(A\subseteq[n]\) of size \(k\), and \(|A|\) for the cardinality of \(A\). In
access-structure statements, complements are always taken inside the player
set \([n]\): for \(A\subseteq[n]\), \(A^c=[n]\setminus A\).

        \subsection{Quantum secret sharing schemes}
A quantum secret sharing (QSS) scheme allows a dealer to distribute a quantum
secret among $n$ players so that certain subsets of players
(called \emph{authorized sets}) can collaborate to reconstruct the secret, while
all other subsets (called \emph{unauthorized sets}) obtain no information about
it. The collection of authorized sets is called the \emph{access structure} of
the scheme.

In this paper, we consider QSS schemes on \emph{qubits}. The secret system $s$
is a single qubit with density matrix $\rho_s\in D(\mathbb{C}^2)$, and each
player also receives a single-qubit share. We assume throughout that a QSS
scheme has at least $3$ players.

\begin{definition}[Quantum secret sharing (QSS) scheme]\label{QSS}
A \emph{quantum secret sharing (QSS) scheme} $\mathcal{S}$ with $n$ players
$(n\geq 3)$ consists of
\begin{itemize}
    \item a completely positive and trace-preserving {\sc Share} map
    \[
        \Lambda_s:\mathcal{B}(\mathcal{H}_s)\rightarrow
        \mathcal{B}\!\left(\bigotimes_{i\in[n]}\mathcal{H}_i\right)
    \]
    that encodes the one-qubit secret state $\rho_s\in D(\mathcal{H}_s)$ into
    a quantum system of $n$ qubits shared among the $n$ players, where
    $\mathcal{H}_s=\mathcal{H}_i=\mathbb{C}^2$. Here $\mathcal{B}(\mathcal{H})$ and $D(\mathcal{H})$ are as defined above;

    \item an access structure $\Gamma\subseteq 2^{[n]}$ and a forbidden
    structure $\varUpsilon\subseteq 2^{[n]}$, whose elements are called
    authorized sets and forbidden sets, respectively;

    \item a set $\{\mathcal{R}_A\}_{A\in\Gamma}$ of completely positive and
    trace-preserving {\sc Reconstruct} maps, where
    \[
        \mathcal{R}_A:
        \mathcal{B}\!\left(\bigotimes_{i\in A}\mathcal{H}_i\right)
        \rightarrow \mathcal{B}(\mathcal{H}_s)
    \]
    recovers the secret state $\rho_s$ from the subsystem held by the players
    in $A$. More precisely, for every $A\in\Gamma$ and every secret state
    $\rho_s\in D(\mathcal{H}_s)$,
    \[
        \mathcal{R}_A\!\left(
        \Tr_{A^c}(\Lambda_s(\rho_s))
        \right)
        =
        \rho_s.
    \]
    Moreover, a subset $B\subseteq[n]$ belongs to the forbidden structure
    $\varUpsilon$ if the reduced density matrix held by $B$ is independent of
    the choice of the secret state, i.e., for any two secret states
    $\rho_s,\sigma_s\in D(\mathcal{H}_s)$,
    \[
        \Tr_{B^c}(\Lambda_s(\rho_s))
        =
        \Tr_{B^c}(\Lambda_s(\sigma_s)).
    \]
\end{itemize}

A QSS scheme is called \emph{pure} if its {\sc Share} map is induced by an
isometry. More precisely, there exists a linear isometry
\[
    V_s:\mathcal{H}_s\rightarrow
    \bigotimes_{i\in[n]}\mathcal{H}_i
\]
such that
\(\Lambda_s(\rho_s)=V_s\rho_s V_s^\dagger\) for every
$\rho_s\in D(\mathcal{H}_s)$. In this case, when the input secret is
a pure state $|\phi\rangle\in\mathcal{H}_s$, we may equivalently write the
encoded state as
\(V_s|\phi\rangle\in\bigotimes_{i\in[n]}\mathcal{H}_i\).

A QSS scheme is called \emph{perfect} if every subset of players is either
authorized or forbidden, namely, \(2^{[n]}=\Gamma\sqcup\varUpsilon\).
\end{definition}
\begin{remark}
Since this paper focuses on perfect QSS schemes, every subset of players is
either authorized or forbidden. Equivalently, every unauthorized set is
forbidden, so the forbidden structure is simply
\(\varUpsilon=2^{[n]}\setminus\Gamma\) and is completely determined by the
access structure \(\Gamma\). It therefore suffices throughout the paper to
specify and study the access structure.

Following the information-rate convention for QSS
schemes~\cite{PhysRevA.83.042324}, a
non-redundant QSS scheme is called \emph{ideal} when it has information rate
one. Equivalently, the entropy of every individual share equals the entropy of
the secret, namely, \(S(i)=S(s)\) for every player \(i\in[n]\).
In the Choi-state representation used here, the one-qubit secret satisfies
\(S(s)=S(R)=\log 2\). Corollary~\ref{coro1} gives \(S(i)=\log 2\) for every
non-redundant player. Hence the pure perfect one-qubit-share schemes with no
redundant players considered in our main results are ideal, and throughout the
paper we refer to this setting as the ideal-qubit setting.
\end{remark}
\subsection{Elementary properties of access structures}

We collect the access-structure properties that will be used throughout the
paper. These follow directly from recoverability, secrecy, perfectness, and
purity, independently of the one-qubit-share assumption.

\begin{proposition}[Basic properties of access structures]
\label{prob}
Let \(\mathcal{S}\) be a pure perfect QSS scheme with access structure
\(\Gamma\). The first two properties below always hold. The third item records
the non-redundancy condition used in this paper.
\begin{enumerate}
    \item \textbf{Monotonicity.} If \(A\in\Gamma\) and
    \(A\subseteq B\subseteq[n]\), then \(B\in\Gamma\).

    \item \textbf{Complementarity.} For every \(A\subseteq[n]\),
    \(A\in\Gamma\) if and only if \(A^c\notin\Gamma\).

    \item \textbf{Non-redundancy.} We say that the scheme has no redundant
    players if, for every player \(i\in[n]\), there exists
    \(A\subseteq[n]\setminus\{i\}\) such that
    \(A\notin\Gamma\) and \(A\cup\{i\}\in\Gamma\).
\end{enumerate}
\end{proposition}

\begin{remark}
The first two properties in Proposition~\ref{prob} are intrinsic to the access
structure of a pure perfect QSS scheme: monotonicity follows from
recoverability, while complementarity follows from purity and perfectness.
The third item is instead our definition of a scheme with no redundant
players and is a natural additional requirement imposed in this paper.
Equivalently, player \(i\) is redundant if no set \(A\) as in item 3 exists;
by monotonicity, adding such a player never changes an unauthorized set into
an authorized one. A redundant player may therefore be removed without
affecting the essential access structure.
\end{remark}

\begin{definition}[Minimal access structure]
The minimal access structure associated with $\Gamma$ is defined as
\[
\Gamma_{\min}
:=
\operatorname{Min}_{\subseteq}(\Gamma)
=
\{A\in \Gamma:\ B\notin \Gamma \text{ for every } B\subsetneq A\}.
\]
Every element of $\Gamma_{\min}$ is called a minimal authorized set.
\end{definition}
By monotonicity, \(\Gamma\) is completely determined by
\(\Gamma_{\min}\): a set is authorized if and only if it contains a member of
\(\Gamma_{\min}\). Moreover, any two authorized sets, and hence any two
members of \(\Gamma_{\min}\), must intersect. Indeed, if authorized sets
\(A\) and \(B\) were disjoint, then \(A\subseteq B^c\), so monotonicity would
make \(B^c\) authorized, contradicting complementarity for \(B\).

\begin{proposition}[Large-set property of minimal access structures]
\label{prob-min}
Let \(\mathcal{S}\) be a pure perfect QSS scheme with minimal access
structure \(\Gamma_{\min}\), and put
\(k:=\min_{A\in\Gamma_{\min}}|A|\). Every subset of \([n]\) of size
\(n-k+1\) contains a member of \(\Gamma_{\min}\).
\end{proposition}

\begin{proof}
Every subset of size \(k-1\) is unauthorized by the definition of \(k\). Its
complement, of size \(n-k+1\), is authorized by Proposition~\ref{prob} and
therefore contains a minimal authorized set.
\end{proof}

\begin{definition}[$k$-Homogeneous and $k$-threshold QSS schemes]
    A QSS scheme $\mathcal{S}$ with minimal access structure $\Gamma_{\min}$ is
    called \emph{$k$-homogeneous} if \(|A|=k\) for every
    \(A\in\Gamma_{\min}\). A $k$-homogeneous QSS scheme is called
    \emph{$k$-threshold} if \(\Gamma_{\min}=\binom{[n]}{k}\).
    Equivalently, a $k$-threshold QSS scheme is one in which a subset of players
    is authorized if and only if it has size at least $k$.
\end{definition}
We give examples of $k$-threshold and $k$-homogeneous QSS schemes. In each
example, we specify the access structure $\Gamma$ and the encoding isometry
$V_s$, but omit the recovery maps $\{\mathcal{R}_A\}_{A\in\Gamma}$.

\begin{example}[$3$-threshold QSS scheme]\label{thres}
    There exists a pure $3$-threshold QSS scheme with $5$ players, obtained
    from the $5$-qubit quantum error-correcting code. Its minimal access
    structure is \(\Gamma_{\min}=\binom{[5]}{3}\).
    Equivalently, every subset of at least three players is authorized, while
    every subset of fewer than three players is forbidden. The explicit encoding
    isometry is given in Appendix~\ref{app:explicit-examples}.
\end{example}

\begin{example}[$3$-homogeneous QSS scheme]\label{homog}
    The Steane code induces a pure $3$-homogeneous QSS scheme with $7$ players.
    Its minimal access structure is
    \[
    \Gamma_{\min}
    =
    \{
    \{1,2,3\},\{1,4,5\},\{1,6,7\},
    \{2,4,6\},\{2,5,7\},\{3,4,7\},\{3,5,6\}
    \}.
    \]
    Every subset of $[7]$ of size $2$ is contained in exactly one minimal
    authorized set of $\Gamma_{\min}$. This combinatorial structure is known
    as the \emph{Fano plane}~\cite{polster1998geometrical}; see
    Figure~\ref{fig:Fano}. The explicit encoding isometry is given in
    Appendix~\ref{app:explicit-examples}.

    \begin{figure}[ht]
        \centering
        \includegraphics[width=3cm]{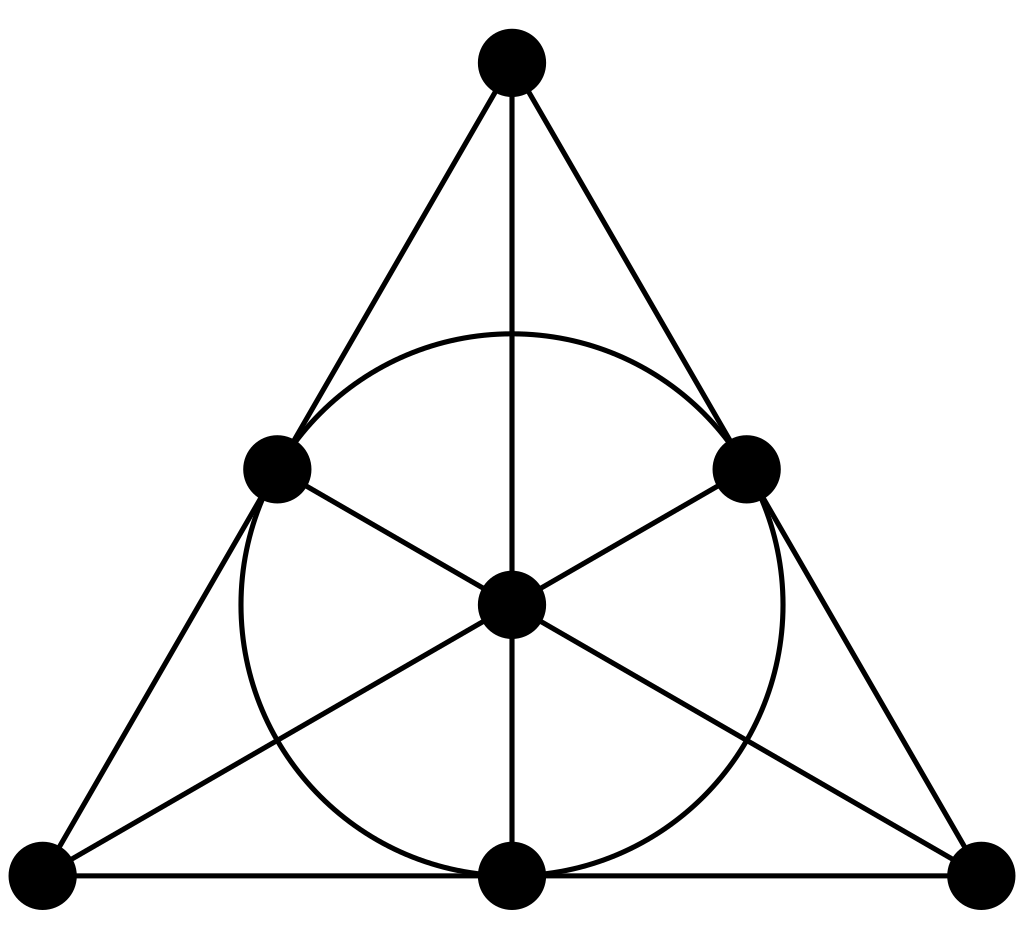}
        \begingroup
        \CLASSOPTIONconferencetrue
        \caption{The Fano plane.}
        \label{fig:Fano}
        \endgroup
    \end{figure}
\end{example}

\subsection{Quantum secret sharing states}
The study of QSS schemes can be viewed as the study of which quantum channels
can satisfy the recoverability and secrecy requirements in Definition~\ref{QSS}.
Indeed, the {\sc Share} map $\Lambda_s$ is a quantum channel from the
secret system to the joint system of all players; in the pure case, it is induced by the isometry $V_s$. The Choi--Jamiolkowski isomorphism allows us to
represent such a channel by a quantum state. More precisely, introduce an
auxiliary reference qubit $R$ and prepare a maximally entangled state between
$R$ and the secret system $s$:
\[
    |\Phi^+\rangle_{RS}
    =
    \frac{1}{\sqrt{2}}\bigl(|00\rangle_{RS}+|11\rangle_{RS}\bigr).
\]
Applying the encoding isometry only to the secret system gives the Choi state
\[
    |\psi_{\mathcal{S}}\rangle
    =
    (\mathbb{I}_R\otimes V_s)|\Phi^+\rangle_{RS}
    =
    \frac{1}{\sqrt{2}}
    \left(
    |0\rangle_R\otimes V_s|0\rangle_S
    +
    |1\rangle_R\otimes V_s|1\rangle_S
    \right).
\]
Thus, the study of the sharing channel can be translated into the study of the
corresponding quantum state. In particular, the recoverability and secrecy
conditions can be expressed in terms of the mutual information between the
reference system $R$ and subsystems of players. This motivates the following
definition of QSS states.

\begin{definition}[Quantum secret sharing (QSS) states]\label{def:qss-state}
Throughout this paper, the term \emph{quantum secret sharing (QSS) state}
refers exclusively to the Choi state associated with a pure perfect QSS
scheme. More precisely, for a pure perfect QSS scheme
$\mathcal{S}=\bigl(\Lambda_s,\Gamma\subseteq 2^{[n]},\{\mathcal{R}_A\}_{A\in\Gamma}\bigr)$
with $n$ players, let $V_s$ be the isometry inducing its {\sc Share} map. The
\emph{quantum secret sharing (QSS) state} associated with $\mathcal{S}$ is
\[
    \ket{\psi_{\mathcal{S}}}
    =
    \frac{1}{\sqrt{2}}
    \bigl(\ket{0}_R\otimes V_s\ket{0}_S+
    \ket{1}_R\otimes V_s\ket{1}_S\bigr)
    \in \mathbb{C}^{2}\otimes(\mathbb{C}^{2})^{\otimes n}.
\]
Thus, in the terminology of this paper, ``QSS state'' always means ``pure
perfect QSS state.'' The Choi state of a pure but non-perfect QSS scheme,
including a ramp scheme with intermediate sets, is not called a QSS state here.
A QSS state is called \emph{$k$-homogeneous} or \emph{$k$-threshold} if the
corresponding pure perfect QSS scheme is $k$-homogeneous or $k$-threshold,
respectively.
\end{definition}
\begin{remark}
Since $\mathcal{S}$ is pure, the QSS state $\ket{\psi_{\mathcal{S}}}$ is a pure
state of $n+1$ qubits indexed by $\{r,1,\ldots,n\}$, where $r$ denotes the
reference system and $\{1,\ldots,n\}$ denotes the players.
Applying the construction above to Examples~\ref{thres} and~\ref{homog}
directly gives the corresponding QSS states, denoted by
$|\psi_{\mathrm{th}}\rangle$ and $|\psi_{\mathrm{St}}\rangle$, respectively;
there is therefore no need to treat them as separate examples. Their explicit
expansions are given in Appendix~\ref{app:explicit-examples}.
\end{remark}
\begin{proposition}[QSS states via mutual information~\cite{imai2003quantuminformationtheoreticalmodel}]\label{QSS-state-MI}
Let \(R:=\{r\}\). Let $|\psi\rangle$ be a pure state of \(n+1\) qubits
indexed by \(R\cup[n]\), where \(R\) is the reference system, and assume that
\(S(R)=\log 2\). Let
$\Gamma\subseteq 2^{[n]}$ be an access structure.
Then $|\psi\rangle$ is a \emph{QSS state with access structure $\Gamma$} in the
sense of Definition~\ref{def:qss-state} if and only if the following two
conditions hold:
\begin{align}
    \textbf{\emph{(Recoverability)}\quad}
    & I(R:A) = 2\log 2,
    && \text{for every authorized set } A\in\Gamma,
    \label{rec}\\
    \textbf{\emph{(Secrecy)}\quad}
    & I(R:B) = 0,
    && \text{for every forbidden set } B\notin\Gamma.
    \label{sec}
\end{align}
Here \(I(R:A)\) denotes the mutual information between the reference system
\(R\) and the subsystem \(A\) in the state $|\psi\rangle$; see
Definition~\ref{mutual}. The condition \(S(R)=\log 2\) states that the
reference qubit is maximally mixed, as it is for the Choi state of a
single-qubit secret.
\end{proposition}
In what follows, we use $R$ to denote the set $\{r\}$.
We collect several entropic facts for QSS states that will be used repeatedly
in the sequel. Throughout this subsection, the secret system is a single qubit,
and $\mathcal{S}$ denotes a pure perfect QSS scheme with access structure
$\Gamma$. Let $|\psi_{\mathcal{S}}\rangle$ be the associated QSS state. For
any subsystem $X\subseteq \{r,1,\ldots,n\}$, we write $S(X)$ for the von
Neumann entropy of the reduced density matrix of $|\psi_{\mathcal{S}}\rangle$
on $X$. Thus, for $A\subseteq[n]$, $S(A)$ always refers to the entropy of the
corresponding subsystem of the QSS state. For explicitly listed player
subsystems, we suppress braces and commas: for example, \(S(12)\) denotes
\(S(\{1,2\})\), and \(S(i)\) denotes \(S(\{i\})\). We retain set notation when
set operations are involved. Since $\mathcal{S}$ is perfect,
every subset $A\notin\Gamma$ is forbidden. For completeness, proofs of the
known facts stated below are provided in Appendix~\ref{app:known-results}.

\begin{theorem}[\cite{imai2003quantuminformationtheoreticalmodel}]
	\label{single}
	Let \(\mathcal S\) be a pure perfect QSS scheme with access structure
	\(\Gamma\). For any unauthorized sets
	\(A,B\notin \Gamma\) such that \(A\cup B\in \Gamma\), we have
	\(S(A),S(B)\geq \log 2\).
\end{theorem}

\begin{corollary}
	\label{coro1}
	Let \(\mathcal S\) be a pure perfect QSS scheme. For any non-redundant
	player \(i\), we have \(S(i)=\log 2\).
\end{corollary}
\begin{lemma}[\cite{helwig2013absolutelymaximallyentangledstates}]
	\label{entropy}
	Let \(\mathcal S\) be a pure perfect QSS scheme with access structure
	\(\Gamma\). If \(B\notin \Gamma\) and there exists some \(i\) such that
	\(B\cup\{i\}\in \Gamma\), then for every subset \(X\subseteq B\),
	\(S(\{i\}\cup X)=S(i)+S(X)\).
\end{lemma}

\begin{corollary}
	\label{min-log}
	Let \(\mathcal S\) be a pure perfect QSS scheme with minimal access
	structure \(\Gamma_{\min}\). For every
	\(A\in\Gamma_{\min}\) and every \(B\subseteq A\), we have
	\(S(B)=|B|\log 2\).
\end{corollary}

We shall also use the following two technical lemmas.

\begin{lemma}[\cite{PhysRevA.111.052410}]
	\label{Zhang}
	Let $|\psi\rangle$ be an $n$-qubit pure state, where $n=4m$ and
	$m\geq 2$. For any $\mathcal{A}\subseteq [n]$ with
	$|\mathcal{A}|=2m+1$, there exists $\mathcal{B}\subseteq \mathcal{A}$ with
	$|\mathcal{B}|=2m$ such that the reduction of $|\psi\rangle$ to
	$\mathcal{B}$ is not maximally mixed.
\end{lemma}

\begin{lemma}[\cite{PhysRevA.83.042324}]
	\label{two}
	Let \(\mathcal S\) be a pure perfect QSS scheme with access structure
	\(\Gamma\). For any \(A,B\in \Gamma\) such that
	\(A\cap B\notin \Gamma\), we have
	\[
	S(A)+S(B)
	\geq
	S(A\cup B)+S(A\cap B)+2\log 2.
	\]
\end{lemma}

We finish the preliminaries with a size bound that will be used in both the
uniformity analysis and the small-player classification.

\begin{lemma}
	\label{kbound}
	Let \(\Gamma_{\min}\subseteq2^{[n]}\) be the minimal access structure of a
	pure perfect QSS scheme with no redundant players, where \(n\geq3\). Put
	\(\ell_{\Gamma_{\min}}:=\min_{A\in\Gamma_{\min}}|A|\). Then
	\[
	2\leq \ell_{\Gamma_{\min}}
	\leq \left\lfloor\frac{n+1}{2}\right\rfloor.
	\]
\end{lemma}

\begin{proof}
	Suppose first that
	\(\ell_{\Gamma_{\min}}>\lfloor(n+1)/2\rfloor\). A set \(A\) of size
	\(\lfloor(n+1)/2\rfloor\) is then unauthorized. By complementarity,
	\(A^c\) is authorized, although
	\[
	|A^c|
	=n-\left\lfloor\frac{n+1}{2}\right\rfloor
	\leq\left\lfloor\frac{n+1}{2}\right\rfloor
	<\ell_{\Gamma_{\min}},
	\]
	a contradiction.
	
	The empty set cannot be authorized, since monotonicity would then make every
	set authorized, contrary to complementarity. If \(\{i\}\) were a minimal
	authorized set, then \([n]\setminus\{i\}\) would be unauthorized. Monotonicity
	would imply that no set avoiding \(i\) is authorized, so no player other than
	\(i\) could occur in a minimal authorized set. Since \(n\geq3\), this
	contradicts the absence of redundant players. Hence
	\(\ell_{\Gamma_{\min}}\geq2\).
\end{proof}
\subsection{$k$-Uniform and absolutely maximally entangled states}

We recall the notions of $k$-uniform and absolutely maximally entangled (AME)
states, and the standard correspondence between AME states and threshold QSS
schemes.

\begin{definition}
    For $k\leq \lfloor n/2\rfloor$, a pure state $|\varphi\rangle$ of $n$
    qudits with density matrix $\rho=\ketbra{\varphi}{\varphi}$ is called a
    \emph{$k$-uniform state} if for every subset $A\subseteq [n]$ with
    $|A|\leq k$, the reduced density matrix
    $\rho_A=\Tr_{A^c}(\rho)$ is maximally mixed, i.e.,
    \(\rho_A=\mathbb{I}_{d^{|A|}}/d^{|A|}\), or equivalently,
    \(S(\rho_A)=|A|\log d\).
    For qubits, $d=2$.

    Furthermore, if $k=\lfloor n/2\rfloor$, then such a $k$-uniform state is
    called an \emph{absolutely maximally entangled (AME) state}, or an
    $\mathrm{AME}(n,d)$ state.
\end{definition}

\begin{remark}
    The QSS state $|\psi_{\mathrm{th}}\rangle$ associated with the scheme in
    Example~\ref{thres} is an $\mathrm{AME}(6,2)$ state. The QSS state
    $|\psi_{\mathrm{St}}\rangle$ associated with the scheme in
    Example~\ref{homog} is a $3$-uniform state of eight qubits.
\end{remark}

The following theorem gives the standard correspondence between threshold QSS
schemes and AME states.

\begin{theorem}[\cite{helwig2013absolutelymaximallyentangledstates}]
\label{AME}
    A state $|\varphi\rangle$ of $2n$ qubits is an $\mathrm{AME}(2n,2)$ state
    if and only if it is a QSS state associated with a pure $n$-threshold QSS
    scheme with $2n-1$ players.
\end{theorem}

\begin{remark}
    The definition of QSS states can be extended to general qudit systems, and
    Theorem~\ref{AME} also holds in the qudit setting.
\end{remark}
The AME--threshold correspondence motivates asking whether homogeneous
non-threshold access structures admit an analogous characterization in terms of
\(k\)-uniform states. We return to this question in
Section~\ref{sec:uniform-qss-comparison}.

\section{QSS States and \(k\)-Uniform States}
\label{sec:uniform-qss-comparison}

Here and throughout the paper, ``QSS state'' refers to the Choi-state notion for
pure perfect schemes specified in Definition~\ref{def:qss-state}.

The AME--threshold correspondence in Theorem~\ref{AME} establishes an exact
correspondence between AME states and threshold QSS states. This raises a
natural question for non-threshold ideal
QSS schemes: to what extent does \(k\)-uniformity capture the QSS-state property
beyond the threshold case?

We approach this question through both positive and negative results. First,
we prove that every pure perfect \(3\)-homogeneous ideal qubit QSS scheme with
no redundant players induces a \(3\)-uniform QSS state. We then give two
examples showing that no unrestricted equivalence holds: a \(k\)-uniform state
need not be a QSS state, and a \(k\)-homogeneous non-threshold QSS scheme need
not induce a \(k\)-uniform state.

\subsection{3-Homogeneous QSS States Imply 3-Uniform States}
\label{sec3}

This subsection proves that pure perfect \(3\)-homogeneous ideal qubit QSS
schemes with no redundant players yield \(3\)-uniform QSS states. By
Lemma~\ref{kbound}, such a scheme contains at least \(5\) players. We analyze
its access structure in Lemma~\ref{x2} and show that the associated QSS state
is \(2\)-uniform in Lemma~\ref{x3}. Lemma~\ref{small-3-hom} settles the small
cases \(n=5,6\), while Lemma~\ref{lem:unauthorized-triples-3hom} proves that
every unauthorized player triple is maximally mixed when \(n\geq7\). These
ingredients are combined in Theorem~\ref{main-lemma} to establish
\(3\)-uniformity.

\begin{lemma}
\label{x2}
Consider a pure perfect $3$-homogeneous QSS scheme $\mathcal{S}$ with minimal
access structure $\Gamma_{\min}\subseteq 2^{[n]}$, where $n\geq 5$. Suppose
that $A_1=\{i,j,k_1\}\in\Gamma_{\min}$ and
$A_2=\{i,j,k_2\}\in\Gamma_{\min}$. Then there exists another minimal authorized
set \(A_3=\{k_1,k_2,k_3\}\in\Gamma_{\min}\), with
$k_3\in [n]\setminus(A_1\cup A_2)$. Moreover, for any minimal authorized set
\(A_3\) of this form,
\(|B\cap(A_1\cup A_2\cup A_3)|\geq 2\) for every \(B\in\Gamma_{\min}\).
\end{lemma}
\begin{proof}
We may assume that $A_1=\{1,2,3\}$ and $A_2=\{1,2,4\}$. Since the scheme is
$3$-homogeneous, all subsets of size $2$ are unauthorized. In particular,
$\{1,2\}$ is unauthorized. By complementarity, its complement $ \{1,2\}^c=\{3,4,\ldots,n\}$ is authorized. Hence it contains some minimal authorized set $A_3\in\Gamma_{\min}$.

Since any two minimal authorized sets must intersect, $A_3$ must intersect both
$A_1$ and $A_2$. Moreover, $A_3\subseteq\{3,4,\ldots,n\}$. Therefore
$A_3\cap A_1\neq\emptyset$ forces $3\in A_3$, and
$A_3\cap A_2\neq\emptyset$ forces $4\in A_3$. Since the scheme is
$3$-homogeneous, $|A_3|=3$, and its third element lies outside
$A_1\cup A_2$. This proves the existence assertion.

To prove the second assertion, fix any minimal authorized set of this form
and, without loss of generality, write it as $A_3=\{3,4,5\}$.

Suppose that there exists $B\in\Gamma_{\min}$ such that
$|B\cap(A_1\cup A_2\cup A_3)|=1$. The case of intersection size $0$ is impossible
by complementarity. Thus $B$ contains a subset $C$ of size $2$ with
$C\subseteq [n]\setminus(A_1\cup A_2\cup A_3)=\{6,7,\ldots,n\}$. Let
$k\in B\cap(A_1\cup A_2\cup A_3)$ be the third element of $B$, so that
$B=C\cup\{k\}$. Then one of $A_1,A_2,A_3$ is disjoint from $B$, contradicting
complementarity. This proves the lemma.
\end{proof}

We next show that $3$-homogeneous QSS states are $3$-uniform. We first
show that they are $2$-uniform.

\begin{lemma}
\label{x3}
Let $\mathcal{S}=\bigl(\Lambda_s,\Gamma,\{\mathcal{R}_A\}_{A\in\Gamma}\bigr)$ be a pure
perfect $3$-homogeneous QSS scheme with $n\geq 5$ players and no redundant
players, and let
$\ket{\psi_{\mathcal{S}}}$ be its associated QSS state. Then
$\ket{\psi_{\mathcal{S}}}$ is $2$-uniform.
\end{lemma}
\begin{proof}
By Corollary~\ref{coro1}, $S(i)=\log 2$ for every $i\in[n]$. Moreover,
$S(R)=\log 2$. Hence $\ket{\psi_{\mathcal{S}}}$ is $1$-uniform.

Now consider an arbitrary subset $A$ of size $2$. If $r\in A$, then
$A\setminus R$ is unauthorized. By Eq.~\eqref{sec},
\(S(A)=S(R)+S(A\setminus R)=2\log 2\).
It remains to consider the case $r\notin A$. Without loss of generality, let
$A=\{1,2\}$. We show that $S(12)=2\log 2$. By non-redundancy, there exists
a minimal authorized set $X$ with $1\in X$. Thus either $|X\cap A|=2$ or
$|X\cap A|=1$.

If $|X\cap A|=2$, then $A\subseteq X$, and Corollary~\ref{min-log} gives
$S(A)=2\log 2$. Otherwise, $|X\cap A|=1$, and we may assume that
    $X^c\cap A=\{2\}$. Since \(X\) is a minimal authorized set,
\(X\setminus\{1\}\) is unauthorized. By complementarity,
\(X^c\cup\{1\}\) is authorized, while \(X^c\) is forbidden. Hence
Lemma~\ref{entropy} applies with \(B=X^c\) and \(i=1\). Strong subadditivity
implies
\[
    S(12)-S(2)
    \geq
    S(X^c\cup\{1\})-S(X^c)
    =\log 2,
\]
where the equality follows from Lemma~\ref{entropy}. On the other hand,
subadditivity gives
\(S(12)\leq S(1)+S(2)=2\log 2\).
Combining these inequalities yields $S(12)=2\log 2$.
\end{proof}
\begin{lemma}[Small cases for $3$-homogeneous QSS schemes]\label{small-3-hom}
Let $\mathcal{S}$ be a pure perfect $3$-homogeneous QSS scheme with \(n\)
players and no redundant players. Then the following hold:
\begin{enumerate}
    \item If $n=5$, then $\mathcal{S}$ is a $3$-threshold QSS scheme. In
    particular, its associated QSS state is an AME$(6,2)$ state, and hence is
    $3$-uniform.
    \item If $n=6$, then no such QSS scheme exists.
\end{enumerate}
\end{lemma}

\begin{proof}
First consider $n=5$. Since the scheme is $3$-homogeneous, every subset of size
$2$ is unauthorized. By complementarity, the complement of every subset of
size $2$ is authorized. Hence every subset of $[5]$ of size $3$ is authorized.
Since all minimal authorized sets have size $3$, we obtain
\(\Gamma_{\min}=\binom{[5]}{3}\).
Thus the scheme is $3$-threshold. By Theorem~\ref{AME}, the associated QSS state
is an AME$(6,2)$ state, and therefore it is $3$-uniform.

Now consider $n=6$. Suppose, for contradiction, that such a pure perfect
$3$-homogeneous QSS scheme exists, and let $\ket{\psi_{\mathcal{S}}}$ be its
associated QSS state. By Lemma~\ref{x3}, this state is $2$-uniform.

In a $3$-homogeneous scheme with six players, a subset $A$ of size $3$ is authorized
if and only if its complement $A^c$ is unauthorized. Hence exactly half of the
subsets of size $3$ are authorized and the other half are unauthorized. In particular,
there exists an unauthorized subset of size $3$; without loss of generality, let
\(A=\{1,2,3\}\) be unauthorized.

For every authorized subset $C$ of size $3$, we have $C\in\Gamma_{\min}$, and hence
Corollary~\ref{min-log} gives \(S(C)=3\log 2\).
For every unauthorized subset $C$ of size $3$, its complement $C^c$ is an
authorized subset of size $3$. Since the global QSS state on $R\cup[6]$ is pure, we have
\(S(C)=S(R\cup C^c)\).
Using the recoverability condition for $C^c$, we obtain
\[
    S(R\cup C^c)
    =
    S(C^c)+S(R)-2\log 2
    =
    2\log 2.
\]
Therefore every unauthorized subset of size $3$ has entropy $2\log 2$.

Since $\{5,6\}$ is unauthorized, its complement $\{1,2,3,4\}$ is authorized.
We distinguish two cases.

If all of
\[
    \{2,3,4\},\quad \{1,3,4\},\quad \{1,2,4\}
\]
are authorized, then $\{3,5,6\}$ and $\{2,5,6\}$ are unauthorized. Strong
subadditivity gives
\[
    S(256)-S(56)
    \ge
    S(2356)-S(356).
\]
Since $\{3,5,6\}$ is unauthorized and $\{2,3,5,6\}$ is authorized,
Lemma~\ref{entropy} gives \(S(2356)-S(356)=\log 2\).
As the QSS state is $2$-uniform, $S(56)=2\log 2$. Hence
\(S(256)\ge 3\log 2\),
contradicting the fact that $\{2,5,6\}$ is an unauthorized subset of size $3$
with entropy $2\log 2$.

Otherwise, at least one of
\[
    \{2,3,4\},\quad \{1,3,4\},\quad \{1,2,4\}
\]
is unauthorized. Without loss of generality, assume that $\{2,3,4\}$ is
unauthorized. Strong subadditivity, applied to the subsystems
\(\{1,2,3\}\) and \(\{2,3,4\}\), gives
\[
    S(123)-S(23)
    \ge
    S(1234)-S(234).
\]
Since the QSS state is \(2\)-uniform,
\(S(23)=S(12)=2\log 2\), so this inequality is equivalently
\[
    S(123)-S(12)
    \ge
    S(1234)-S(234).
\]
Since $\{2,3,4\}$ is unauthorized and $\{1,2,3,4\}$ is authorized,
Lemma~\ref{entropy} gives \(S(1234)-S(234)=\log 2\).
We therefore obtain \(S(123)\ge 3\log 2\),
contradicting the fact that $\{1,2,3\}$ is an unauthorized subset of size $3$
with entropy $2\log 2$.

Both cases lead to contradictions. Therefore no pure perfect $3$-homogeneous
QSS scheme with six players exists.
\end{proof}

\begin{lemma}[Unauthorized triples]\label{lem:unauthorized-triples-3hom}
Let \(\mathcal{S}\) be a pure perfect \(3\)-homogeneous QSS scheme with
\(n\geq 7\) players and no redundant players, and let
\(\ket{\psi_{\mathcal{S}}}\) be its associated QSS state. If
\(A\in\binom{[n]}{3}\) is unauthorized, then
\(S(A)=3\log 2\).
\end{lemma}

\begin{proof}
By Lemma~\ref{x3}, the QSS state is \(2\)-uniform. Without loss of generality,
let \(A=\{1,2,3\}\). For \(i=1,2,3\), define
\(\mathcal{X}_i=\{X\in\Gamma_{\min}: i\in X\}\).
By non-redundancy, each \(\mathcal X_i\) is nonempty.

We first consider the case where there exists
\(X\in\mathcal X_1\cup\mathcal X_2\cup\mathcal X_3\) such that
\(|X\cap A|=1\). Strong subadditivity and Lemma~\ref{entropy} give
\[
    S(A)-S(X^c\cap A)\geq S(X^c\cup A)-S(X^c)=\log 2.
\]
Since the state is \(2\)-uniform and \(X^c\cap A\) has size \(2\), this gives
\(S(A)\geq 3\log 2\). Subadditivity gives \(S(A)\leq 3\log 2\), and hence
\(S(A)=3\log 2\).

It remains to consider the case where \(|X\cap A|=2\) for every
\(X\in\mathcal X_1\cup\mathcal X_2\cup\mathcal X_3\). Assume
\(\{1,2,4\}\in\Gamma_{\min}\). Then \(\{1,2,3,4\}\) is authorized. If there is
a set \(B\in\binom{\{1,2,3\}}{2}\) such that \(B\cup\{4\}\) is unauthorized,
then strong subadditivity and Lemma~\ref{entropy} give
\[
    S(A)-S(B)\geq S(1234)-S(B\cup\{4\})=\log 2.
\]
Since the state is \(2\)-uniform, \(S(B)=2\log 2\), and again
\(S(A)=3\log 2\). Therefore we may assume that every set
    \(\{4\}\cup B\), with \(B\in\binom{\{1,2,3\}}{2}\), is a minimal
    authorized set. In
particular,
\[
    \{1,2,4\},\quad \{1,3,4\},\quad \{2,3,4\}
\]
are minimal authorized sets.

Apply Lemma~\ref{x2} to the minimal authorized sets \(\{1,2,4\}\) and
\(\{1,3,4\}\). After relabeling the players outside \(\{1,2,3,4\}\), denote the third
minimal authorized set furnished by the lemma by \(\{2,3,5\}\). Similarly,
applications of Lemma~\ref{x2} to
\(\{1,2,4\},\{2,3,4\}\) and to
\(\{1,3,4\},\{2,3,4\}\), respectively, furnish minimal authorized sets
\(\{1,3,x\}\) and \(\{1,2,y\}\) for some \(x,y\in\{5,\ldots,n\}\).

Suppose first that at least one of \(x,y\) differs from \(5\). After possibly
interchanging the labels \(2\) and \(3\), and then relabeling the players
outside \(\{1,2,3,4,5\}\), we may assume that \(x=6\). Then the authorized sets
include
\[
    \{1,2,4\},\quad \{1,3,4\},\quad \{2,3,4\},\quad
    \{2,3,5\},\quad \{1,3,6\}.
\]
Since \(A=\{1,2,3\}\) is unauthorized, its complement
\(\{4,\ldots,n\}\) is authorized and contains a minimal authorized subset
\(S\subseteq\{4,\ldots,n\}\). Since any two authorized sets must intersect,
\(S\) must intersect all the displayed authorized triples, forcing
\(S=\{4,5,6\}\). Thus the minimal authorized sets include
\[
    \{1,2,4\},\quad \{1,3,4\},\quad \{2,3,4\},\quad
    \{2,3,5\},\quad \{1,3,6\},\quad \{4,5,6\}.
\]

We next record two consequences of Lemma~\ref{x2}. Apply it to
\(\{1,2,4\}\) and \(\{1,3,4\}\), taking \(\{2,3,5\}\) as the third minimal
authorized set supplied by the lemma. Every minimal authorized set must then
contain at least two elements of \(\{1,2,3,4,5\}\), so no minimal authorized
set can contain \(\{6,7\}\). Similarly, apply Lemma~\ref{x2} to
\(\{1,2,4\}\) and \(\{2,3,4\}\), taking \(\{1,3,6\}\) as the third set. It
follows that no minimal authorized set can contain \(\{5,7\}\).

Let \(A_7\) be a minimal authorized set containing \(7\), whose existence
follows from non-redundancy. Since \(A_7\) must intersect \(\{4,5,6\}\), while
it cannot contain either \(5\) or \(6\), we have
\(A_7=\{4,7,z\}\) for some \(z\). The set \(A_7\) must intersect both
\(\{2,3,5\}\) and \(\{1,3,6\}\). Since \(z\notin\{5,6\}\), these two
intersection requirements give
\[
    z\in\{2,3\}\cap\{1,3\}=\{3\}.
\]
Thus \(A_7=\{3,4,7\}\). On the other hand, the minimal authorized set
\(\{1,2,y\}\) must intersect \(\{4,5,6\}\). Since \(y\geq 5\), this forces
\(y\in\{5,6\}\). Consequently,
\[
    \{3,4,7\}\cap\{1,2,y\}=\emptyset,
\]
contradicting the intersecting property of authorized sets.

It remains to consider the case \(x=y=5\). Then the minimal authorized sets
include
\[
    \{1,2,4\},\quad \{1,3,4\},\quad \{2,3,4\},\quad
    \{2,3,5\},\quad \{1,2,5\},\quad \{1,3,5\}.
\]
As above, the complement \(\{4,\ldots,n\}\) of the unauthorized set
\(A=\{1,2,3\}\) is authorized and contains a minimal authorized set \(S\).
Its intersections with the first three displayed triples force \(4\in S\),
while its intersections with the last three force \(5\in S\). Relabeling the
third element of \(S\), we may write \(S=\{4,5,6\}\).

Applying Lemma~\ref{x2} gives the following four restrictions:
\begin{enumerate}
    \item no subset of size \(2\) drawn from \(\{3,7,8,\ldots,n\}\) is contained in a minimal
    authorized set;
    \item no subset of size \(2\) drawn from \(\{2,7,8,\ldots,n\}\) is contained in a minimal
    authorized set;
    \item no subset of size \(2\) drawn from \(\{1,7,8,\ldots,n\}\) is contained in a minimal
    authorized set;
    \item no subset of size \(2\) drawn from \(\{6,7,8,\ldots,n\}\) is contained in a minimal
    authorized set.
\end{enumerate}

Indeed, the first three restrictions follow from the second assertion of
Lemma~\ref{x2} by taking, respectively, the following triples as
\((A_1,A_2,A_3)\):
\[
    \{1,2,4\},\{1,2,5\},\{4,5,6\};\qquad
    \{1,3,4\},\{1,3,5\},\{4,5,6\};\qquad
    \{2,3,4\},\{2,3,5\},\{4,5,6\}.
\]
For the fourth restriction, use the second assertion of Lemma~\ref{x2} with
\(\{1,2,4\}\), \(\{1,3,4\}\), and \(\{2,3,5\}\) as
\((A_1,A_2,A_3)\). In each application, the complement of the union of the
three displayed triples is exactly the set appearing in the corresponding
restriction, and Lemma~\ref{x2} rules out a minimal authorized set containing
two elements of that complement.

Let \(A_7\) be a minimal authorized set containing \(7\). The four restrictions
exclude every other element except \(4\) and \(5\), and hence
\(A_7=\{4,5,7\}\).

Finally, \(\{4,5\}\) is unauthorized, so its complement
\[
    H:=\{1,2,3,6,7,\ldots,n\}
\]
is authorized and contains a minimal authorized set \(P\subseteq H\).
The set \(P\) must intersect \(\{4,5,6\}\), and since
\(P\cap\{4,5\}=\emptyset\), this forces \(6\in P\). Similarly, \(P\) must
intersect \(\{4,5,7\}\), which forces \(7\in P\). Thus
\(\{6,7\}\subseteq P\), contradicting the fourth restriction. This proves the
lemma.
\end{proof}

\begin{theorem}
\label{main-lemma}
Let $\mathcal{S}=\bigl(\Lambda_s,\Gamma,\{\mathcal{R}_A\}_{A\in\Gamma}\bigr)$ be a pure
perfect $3$-homogeneous QSS scheme with \(n\) players and no redundant players.
Then its associated QSS state $\ket{\psi_{\mathcal{S}}}$ is $3$-uniform.
\end{theorem}

\begin{proof}
By Lemma~\ref{x3}, the QSS state is already \(2\)-uniform. We prove that
\(S(A)=3\log 2\) for every subsystem \(A\) of size \(3\).

The case \(n=5\) follows from Lemma~\ref{small-3-hom}, and the case \(n=6\) is
impossible by the same lemma. Hence, in the rest of the proof, we may assume
that \(n\ge 7\).

Let \(A\) be a subsystem of size \(3\). If \(r\in A\), then \(A\setminus R\)
has size \(2\) and is unauthorized. By secrecy and \(2\)-uniformity,
\(S(A)=S(R)+S(A\setminus R)=3\log 2\).
If \(r\notin A\) and \(A\) is authorized, then \(A\) is a minimal authorized
set, so
Corollary~\ref{min-log} gives \(S(A)=3\log 2\). Finally, if
\(r\notin A\) and \(A\) is unauthorized, Lemma~\ref{lem:unauthorized-triples-3hom}
gives \(S(A)=3\log 2\). Hence every three-qubit reduction is maximally mixed,
and the QSS state is \(3\)-uniform.
\end{proof}

\subsection{Limitations Beyond the Threshold Case}

The positive result above suggests two possible converses. First, given a
\(k\)-uniform state, can one designate one of its qubits as a reference system
so that the remaining qubits form a perfect QSS scheme? Second, if an ideal QSS
scheme has a \(k\)-homogeneous minimal access structure, must its associated
QSS state be \(k\)-uniform? The following examples answer both questions
negatively, for two different structural reasons.

\subsubsection{\(k\)-uniformity does not imply the QSS-state property}

Uniformity constrains small reductions without distinguishing any subsystem as
the reference. By contrast, after a reference qubit \(R\) is chosen, a perfect
QSS state must satisfy the reference-dependent dichotomy
\(I(R:A)\in\{0,2\log 2\}\) for every player set \(A\). Thus a genuine
counterexample must rule out every possible choice of \(R\), not merely one
particular choice. The next example has exactly this property.

\begin{example}[No reference qubit turns a \(3\)-uniform state into a QSS state]
\label{ex:uniform-not-qss}
Zhang et al.\ constructed the eight-qubit graph state \(|T_4\rangle\) in
Sec.~V-A of~\cite{PhysRevA.111.052410}. Their state has the extremal number,
\(56\), of maximally mixed four-qubit reductions, and they also proved that it
is \(3\)-uniform. The adjacency matrix of their graph \(T_4\) is
\[
A_{T_4}=
\begin{bmatrix}
0&1&1&1&1&0&0&0\\
1&0&1&1&0&1&0&0\\
1&1&0&1&0&0&1&0\\
1&1&1&0&0&0&0&1\\
1&0&0&0&0&1&1&1\\
0&1&0&0&1&0&1&1\\
0&0&1&0&1&1&0&1\\
0&0&0&1&1&1&1&0
\end{bmatrix}.
\]
Recall that for a graph state \(|G\rangle\) with adjacency matrix \(A_G\),
the entropy of the reduction to a vertex set \(X\) is
\[
S(X)=\rank_{\mathbb F_2}(A_{X,X^c})\log 2,
\]
where \(A_{X,X^c}\) is the submatrix with rows indexed by \(X\) and columns
indexed by \(X^c\)~\cite{Hein_2004}. For the matrix above, a direct rank
calculation confirms that
\(\rank_{\mathbb F_2}(A_{X,X^c})=|X|\) for every \(|X|\leq 3\). Hence
\(|T_4\rangle\) is \(3\)-uniform. On the other hand, for
\(K=\{1,2,3,8\}\), we have \(\rank_{\mathbb F_2}(A_{K,K^c})=3\), and therefore
\(S(K)=3\log 2\).

We now show that no designation of a reference qubit produces a QSS state.
Choose an arbitrary qubit as the putative reference system \(R\), and let
\[
L=
\begin{cases}
K, & R\in K,\\
K^c, & R\notin K.
\end{cases}
\]
Then \(R\in L\), \(|L|=4\), and purity gives
\(S(L)=S(K)=S(K^c)=3\log 2\). Put \(A=L\setminus R\). Since
\(|T_4\rangle\) is \(3\)-uniform,
\(S(R)=\log 2\) and \(S(A)=3\log 2\). Consequently,
\[
I(R:A)=S(R)+S(A)-S(L)=\log 2.
\]

Thus every possible reference qubit admits a corresponding player set \(A\)
for which \(I(R:A)\) is neither \(0\) nor \(2\log 2\). By
Proposition~\ref{QSS-state-MI}, no choice of a qubit as the reference system can
turn \(|T_4\rangle\) into a pure perfect QSS state. This conclusion concerns
the notion of a QSS state fixed in Definition~\ref{def:qss-state}; it does not
exclude a non-perfect or ramp interpretation.
\end{example}

\subsubsection{\(k\)-homogeneity does not lead to \(k\)-uniformity}

The converse failure has a different source. For a binary self-dual code
\(C\subseteq\mathbb F_2^N\), consider the code state
\[
|\Phi_C\rangle=|C|^{-1/2}\sum_{c\in C}|c\rangle
\]
and, for \(X\subseteq[N]\), the shortened subcode
\[
C_X=\{c\in C:\supp(c)\subseteq X\}.
\]
The stabilizer entropy formula~\cite{7918542} gives
\[
S(X)=\bigl(|X|-2\dim C_X\bigr)\log 2.
\]
This formula separates the two relevant mechanisms. After a reference
coordinate \(R\) is fixed, the dealer-relative inclusion-minimal codeword
supports containing \(R\) determine the minimal access structure. In contrast,
the global minimum distance \(d(C)\) determines the uniformity of the code
state: \(|\Phi_C\rangle\) is \((d(C)-1)\)-uniform, while a codeword of weight
\(d(C)\) prevents \(d(C)\)-uniformity. These two code parameters need not agree.
The next example exploits precisely this separation.

\begin{example}[A \(5\)-homogeneous QSS state that is only \(3\)-uniform]
\label{ex:k-uniform-counterexample}
We give a concrete self-dual \([18,9,4]\) code for which the two parameters
above separate. Let \(C\subseteq\mathbb F_2^{18}\) be generated by the
following matrix, whose rows are written as binary strings in coordinate order
\(1,\ldots,18\):
\[
G=
\left[
\begin{array}{c}
\mathtt{100000000000100011}\\
\mathtt{010000000111101110}\\
\mathtt{001000000111001111}\\
\mathtt{000100000101011100}\\
\mathtt{000010000110011100}\\
\mathtt{000001000011011100}\\
\mathtt{000000100111101101}\\
\mathtt{000000010111011000}\\
\mathtt{000000001111010100}
\end{array}
\right].
\]
This matrix has the form \(G=[\mathbb{I}_9\mid A]\), and a direct calculation gives
\(AA^T=\mathbb{I}_9\) over \(\mathbb F_2\). Hence \(GG^T=0\). Since
\(\rank(G)=9=18/2\), the code is self-dual, i.e., \(C=C^\perp\).
The weight enumerator of \(C\), obtained by enumerating its \(2^9\) codewords,
is
\[
    W_C(y)=
    1+17y^4+51y^6+187y^8+187y^{10}
    +51y^{12}+17y^{14}+y^{18}.
\]
In particular, \(d(C)=4\).

Choose coordinate \(14\) as the dealer/reference coordinate \(R\), and regard
the other \(17\) coordinates as players. Since \(d(C)=4\), we have \(C_R=0\)
and hence \(S(R)=\log 2\). For any player set \(A\), the entropy formula gives
\(I(R:A)=2(\dim C_{R\cup A}-\dim C_A)\log 2\). Adding the single coordinate
\(R\) can increase the subcode dimension by at most one, so
\(I(R:A)\in\{0,2\log 2\}\). Define
\[
    \Gamma=\{A:\dim C_{R\cup A}-\dim C_A=1\}.
\]
Then \(|\Phi_C\rangle\) is a pure perfect QSS state with access structure
\(\Gamma\) by Proposition~\ref{QSS-state-MI}. The condition
\(\dim C_{R\cup A}-\dim C_A=1\) is equivalent to the existence of a codeword
\(c\in C\) such that \(R\in\supp(c)\subseteq A\cup R\). Thus the minimal
authorized sets are precisely the dealer-relative inclusion-minimal supports of
codewords containing \(R\), with \(R\) removed, in agreement with the
general self-dual-code construction of Sarvepalli and
Raussendorf~\cite[Corollary~5]{PhysRevA.81.052333}.

We now evaluate the two controlling parameters separately. First, direct
enumeration of the \(2^9\) codewords, reproduced in
Appendix~\ref{app:self-dual-code-verification}, shows that all dealer-relative
inclusion-minimal codeword supports containing \(R=14\) have weight \(6\):
there are exactly \(51\) of them. After deleting \(R\), they give \(51\)
minimal authorized sets of size \(5\). Hence \(\Gamma_{\min}\) has profile
\(5^{51}\), so the QSS scheme is \(5\)-homogeneous but not \(5\)-threshold,
since \(51<\binom{17}{5}\).

The scheme has no redundant players. Indeed, the following five minimal
codeword supports containing \(R=14\) cover all non-reference coordinates:
\[
    \{4,10,12,14,15,16\},\quad
    \{2,4,11,13,14,17\},\quad
    \{4,5,6,8,9,14\},
\]
\[
    \{1,2,4,11,14,18\},\quad
    \{2,3,4,7,11,14\}.
\]
Deleting coordinate \(14\) gives minimal authorized sets whose union contains
every player.

Second, the global minimum distance is only \(d(C)=4\). Therefore the code state
is \(3\)-uniform. It is not \(4\)-uniform: the first generator row is a
weight-four codeword with support \(B=\{1,13,17,18\}\), which avoids the
reference coordinate \(R=14\). Thus \(\dim C_B=1\), and
\[
S(B)=(|B|-2\dim C_B)\log 2=2\log 2<4\log 2.
\]
The reduction to \(B\) is therefore not maximally mixed. In particular, the
associated QSS state cannot be \(5\)-uniform, even though its minimal access
structure is \(5\)-homogeneous. The example separates the dealer-relative
parameter controlling \(\Gamma_{\min}\) from the global distance controlling
uniformity.
\end{example}

The two examples above show that the close correspondence between uniformity
and QSS breaks down outside the threshold setting. A \(k\)-uniform state need not
determine any perfect QSS access structure, and a \(k\)-homogeneous
non-threshold ideal QSS scheme need not induce a \(k\)-uniform QSS state.

	\section{Pure QSS Schemes with Small Numbers of Players}
	
	In this section, we classify realizable minimal access structures of pure
perfect QSS schemes with no redundant players and at most seven players in
the one-qubit-share setting.

The following theorem is the main classification theorem of this section. Its
proof is obtained by combining the homogeneous classification in
Subsection~\ref{4.1} with the non-homogeneous analysis in
Subsection~\ref{gas}.

\begin{theorem}[Classification for at most seven players]\label{thm:classification-n-leq-7}
Let $\mathcal{S}$ be a pure perfect QSS scheme that shares a single-qubit secret
among $n$ players, with each player receiving one qubit. Assume that the scheme
has no redundant players. If $3\le n\le 7$, then such a QSS scheme
exists if and only if $n=5$ or $n=7$. More precisely:
\begin{enumerate}
    \item If $n=5$, then, up to relabeling of players, the only possible
    minimal access structure is the $3$-threshold access structure
    \(\Gamma_{\min}=\binom{[5]}{3}\),
    which is realized by the QSS scheme in Example~\ref{thres}.

    \item If $n=7$, then, up to relabeling of players, the only possible
    minimal access structure is the Fano-plane access structure
    \[
    \Gamma_{\min}
    =
    \bigl\{
    \{1,2,3\},\{1,4,5\},\{1,6,7\},
    \{2,4,6\},\{2,5,7\},\{3,4,7\},\{3,5,6\}
    \bigr\},
    \]
    which is realized by the Steane-code QSS scheme in Example~\ref{homog}.
\end{enumerate}
Here and throughout this section, relabeling acts only on the player labels;
the dealer/reference system used to form the QSS state is kept fixed.
In particular, no such QSS scheme exists for $n=3,4$, or $6$.
\end{theorem}
    \subsection{Classification of homogeneous minimal access structures for $n\leq 7$}\label{4.1}
    
    In this subsection, we classify the realizable homogeneous minimal access
structures of pure perfect one-qubit-share QSS schemes with no redundant
players and with at most \(7\) players.

We first establish the homogeneous classification for all \(3\leq n\leq 7\).
The following lemma treats the extremal odd-player case in which the smallest
minimal authorized sets have the largest size permitted by
Lemma~\ref{kbound}; it will identify the corresponding cases in the proof of
Theorem~\ref{mainthm} as threshold schemes.

\begin{lemma}
\label{tight}
Let \(\mathcal{S}\) be a pure perfect QSS scheme with minimal access structure
\(\Gamma_{\min}\subseteq2^{[n]}\), where \(n\geq3\). If \(n\) is odd and
\(\ell_{\Gamma_{\min}}=(n+1)/2\), then \(\mathcal{S}\) is an
\(\ell_{\Gamma_{\min}}\)-threshold QSS scheme.
\end{lemma}

\begin{proof}
Every set of size \(\ell_{\Gamma_{\min}}-1\) is unauthorized. By
complementarity, its complement, whose size is
\[
    n-(\ell_{\Gamma_{\min}}-1)
    =\frac{n+1}{2}
    =\ell_{\Gamma_{\min}},
\]
is authorized. Hence every subset of size \(\ell_{\Gamma_{\min}}\) is authorized,
which is exactly the threshold property.
\end{proof}

The next lemma rules out a competing family of homogeneous access structures.
In the proof of Theorem~\ref{mainthm}, it is combined with non-redundancy to
exclude \(2\)-homogeneous schemes for the relevant numbers of players.

\begin{lemma}
\label{notwo}
Let \(\mathcal{S}\) be a pure perfect QSS scheme with \(n\geq4\) players and
no redundant players. Then its minimal access structure \(\Gamma_{\min}\)
contains at most one minimal authorized set of size \(2\).
\end{lemma}

\begin{proof}
Assume for contradiction that \(\Gamma_{\min}\) contains two distinct minimal
authorized sets of size \(2\). Any two authorized sets intersect, so, after
relabeling, we may take
\(\{1,2\},\{1,3\}\in\Gamma_{\min}\). The singleton \(\{1\}\) is
unauthorized. Lemma~\ref{two} gives
\[
    S(12)+S(13)\geq S(123)+S(1)+2\log2.
\]
By Corollaries~\ref{min-log} and~\ref{coro1},
\(S(12)=S(13)=2\log2\) and \(S(1)=\log2\), so
\(S(123)\leq\log2\). On the other hand, \(\{1,2,3\}\) is authorized, and
recoverability implies
\(S(R123)=S(123)-\log2\). Nonnegativity therefore gives
\(S(123)\geq\log2\). Consequently, \(S(R123)=0\).

The global QSS state thus factorizes across
\[
    R123\,|\,\{4,\ldots,n\}.
\]
For any player set \(A\), put \(A_0=A\cap\{1,2,3\}\). The factorization
implies \(I(R:A)=I(R:A_0)\). Hence, if \(A\) is authorized, then \(A_0\) is
also authorized by the mutual-information characterization of perfect QSS
states. If a minimal authorized set contained a player from
\(\{4,\ldots,n\}\), its intersection with \(\{1,2,3\}\) would be a proper
authorized subset, contradicting minimality. Thus no such player occurs in a
minimal authorized set, contrary to non-redundancy.
\end{proof}

\begin{samepage}
\begin{theorem}[Homogeneous classification for at most seven players]
\label{mainthm}
Let \(\mathcal S\) be a pure perfect \(k\)-homogeneous one-qubit-share QSS
scheme with \(3\leq n\leq 7\) players and no redundant players. Then, up to
relabeling of the players, exactly the following homogeneous minimal access
structures are realizable:
\begin{enumerate}
    \item If \(n=5\), then necessarily \(k=3\) and
    \[
        \Gamma_{\min}=\binom{[5]}{3},
    \]
    the \(3\)-threshold minimal access structure. It is realized by the QSS
    scheme in Example~\ref{thres}.
    \item If \(n=7\), then necessarily \(k=3\) and
    \[
    \Gamma_{\min}
    =
    \bigl\{
    \{1,2,3\},\{1,4,5\},\{1,6,7\},
    \{2,4,6\},\{2,5,7\},\{3,4,7\},\{3,5,6\}
    \bigr\},
    \]
    the Fano-plane minimal access structure. It is realized by the
    Steane-code QSS scheme in Example~\ref{homog}.
\end{enumerate}
No such homogeneous QSS scheme exists for \(n=3,4\), or \(6\).
\end{theorem}
\end{samepage}

\begin{proof}
By Lemma~\ref{kbound}, we have
\(2\le k\le \left\lfloor (n+1)/2\right\rfloor\). We analyze the cases
$3\le n\le 7$.

\textbf{Case $n=3$.}
The only possible value is $k=2$. Every singleton is unauthorized, so
complementarity makes every pair authorized. Thus the scheme would be
$2$-threshold and, equivalently, its QSS state would be an AME$(4,2)$ state by
Theorem~\ref{AME}. Since no AME$(4,2)$ state exists~\cite{Huber_2018}, this
case is impossible.

\textbf{Case $n=4$.}
By Lemma~\ref{kbound}, we have $k=2$. However, Lemma~\ref{notwo}
implies that $\Gamma_{\min}$ contains at most one minimal authorized set
of size $2$, contradicting non-redundancy. Hence no pure perfect
$k$-homogeneous QSS scheme exists for $n=4$.

\textbf{Case $n=5$.}
By Lemma~\ref{kbound}, we have $2\le k\le 3$. The case $k=2$ is ruled
out by Lemma~\ref{notwo}, together with non-redundancy: a
$2$-homogeneous scheme would need minimal authorized pairs covering all
players, whereas Lemma~\ref{notwo} allows at most one such pair. Hence
$k=3$. By Lemma~\ref{tight}, the scheme must be $3$-threshold, and this
minimal access structure is realized by Example~\ref{thres}.

\textbf{Case $n=6$.}
By Lemma~\ref{kbound}, we have $k\in\{2,3\}$. The case $k=2$ is ruled
out by the same argument as in the case $n=4$. The case $k=3$ is ruled
out by Lemma~\ref{small-3-hom}. Hence no pure perfect $k$-homogeneous QSS scheme
exists for $n=6$.

\textbf{Case $n=7$.}
By Lemma~\ref{kbound}, the possible values are $2\le k\le 4$.
The case $k=2$ is ruled out by Lemma~\ref{notwo} and non-redundancy, as
above. If $k=4$, then Lemma~\ref{tight} implies that the scheme is
$4$-threshold, which would require an AME$(8,2)$ state by
Theorem~\ref{AME}. Since no AME$(8,2)$ state exists, this case is also
impossible. It remains to classify the case $k=3$.

Let $\rho$ denote the density matrix of the associated QSS state, and let
$\Gamma_{\min}=\{A_1,A_2,\ldots,A_s\}$ be the minimal access structure. By
Theorem~\ref{main-lemma}, $\rho$ is a $3$-uniform state, so
$S(X)=3\log 2$ for every three-qubit subsystem $X$.

Consider any subset $A\subseteq T=\{r,1,\ldots,7\}$ of size $4$. Since the
global QSS state is pure, the Schmidt decomposition implies that $\rho_A$ is
maximally mixed if and only if $\rho_{T\setminus A}$ is maximally mixed.
Combining this observation with Eqs.~\eqref{rec} and~\eqref{sec} gives the
following two facts:
\begin{enumerate}
    \item If $r\in A$, then $A\setminus R$ is authorized if and only if
    $\rho_A\neq 2^{-4}\mathbb{I}$. Indeed, $S(A\setminus R)=3\log 2$. If
    $A\setminus R$ is authorized, recoverability gives $S(A)=2\log 2$; if it
    is forbidden, secrecy gives $S(A)=4\log 2$.
    \item If $r\notin A$, then $A$ is unauthorized if and only if
    $\rho_A\neq 2^{-4}\mathbb{I}$. Indeed, the complement of $R\cup A$ has three
    qubits and hence entropy $3\log 2$. If $A$ is unauthorized, secrecy and
    purity give $S(A)=2\log 2$; if it is authorized, recoverability and purity
    give $S(A)=4\log 2$.
\end{enumerate}

Suppose that a pair $X$ is not contained in any minimal authorized set.
Without loss of generality, let $X=\{1,2\}$. Then every triple
$\{1,2,i\}$ with $i\in\{3,4,5,6,7\}$ is unauthorized. By complementarity,
$\{3,4,5,6,7\}\setminus\{i\}$ is authorized. The second fact above therefore
shows that every subset of size $4$ contained in $\{3,4,5,6,7\}$ has maximally mixed
reduction, contradicting Lemma~\ref{Zhang}. Hence every pair is contained in
at least one member of $\Gamma_{\min}$.

We next prove that no pair is contained in two distinct members of
$\Gamma_{\min}$. Suppose otherwise and, after relabeling, take
$A_1=\{1,2,3\}$ and $A_2=\{1,2,4\}$ in $\Gamma_{\min}$. By
Lemma~\ref{x2}, after relabeling the players outside $A_1\cup A_2$, there is a minimal
authorized set $A_3=\{3,4,5\}$ such that every
$B\in\Gamma_{\min}$ satisfies
\[
    |B\cap(A_1\cup A_2\cup A_3)|\geq2.
\]
The pair $\{6,7\}$ must be contained in some $B\in\Gamma_{\min}$ by the
preceding paragraph. Since $|B|=3$, however, it contains only one element of
$A_1\cup A_2\cup A_3=\{1,2,3,4,5\}$, a contradiction. Thus every pair is
contained in at most one member of $\Gamma_{\min}$.

Consequently, every pair of players is contained in exactly one minimal
authorized triple. Counting incidences between pairs and triples gives
$3|\Gamma_{\min}|=\binom{7}{2}=21$, so $|\Gamma_{\min}|=7$. Therefore
$\Gamma_{\min}$ is a Steiner triple system $S(2,3,7)$, which is unique up to
isomorphism. Hence, after relabeling the players,
\[
\Gamma_{\min}
=
\bigl\{
\{1,2,3\},\{1,4,5\},\{1,6,7\},
\{2,4,6\},\{2,5,7\},\{3,4,7\},\{3,5,6\}
\bigr\}.
\]
Thus the unique minimal access structure in the seven-player
$3$-homogeneous case is the Fano plane. Example~\ref{homog} shows that this
minimal access structure is realizable via the Steane-code construction.

Combining all cases proves the stated classification.
\end{proof}

	\subsection{The non-homogeneous cases for $n\leq 7$}\label{gas}
	Having classified homogeneous minimal access structures in Subsection~\ref{4.1}, it remains to handle
the non-homogeneous cases. This subsection proves that non-homogeneous access structures
do not give rise to any additional pure perfect one-qubit-share QSS schemes for
\(n\leq 7\) with no redundant players. Together with the homogeneous classification, this completes the proof of
Theorem~\ref{thm:classification-n-leq-7}.

In each non-homogeneous case considered below, we first show that some pair of
players is a minimal authorized set. Lemma~\ref{notwo} implies that there is
only one such pair; after relabeling the players, we write it as
$\{1,2\}$. Every other minimal authorized set must contain at least one of
the players $1$ and $2$, because any two authorized sets intersect. It cannot
contain both, since it would then properly contain the already authorized set
$\{1,2\}$ and would not be minimal. Hence every other minimal authorized set
contains exactly one of these two players.

In the cases below, all the remaining minimal authorized sets are triples.
After removing player $1$ or player $2$ from such a triple, the other two
players can be viewed as an edge. This gives two graphs: one records the
triples containing player $1$, and the other records the triples containing
player $2$. The next lemma describes the relation between these two graphs. We
use it in Theorem~\ref{general} for the six-player case and again in
Theorem~\ref{thm:no-nonhomogeneous-n7} for the seven-player case.

For a graph $G=(U,E)$, a set $C\subseteq U$ is called a
\emph{vertex cover} if
\[
C\cap e\neq\varnothing
\qquad
\text{for every } e\in E.
\]
A vertex cover is \emph{inclusion-minimal} if none of its proper
subsets is a vertex cover.

\begin{lemma}[Graph reduction around a minimal authorized pair]
    \label{lem:minimal-pair-graphs}
    Let $\mathcal{S}$ be a pure perfect QSS scheme with minimal access
    structure $\Gamma_{\min}$. Suppose that
    $\{1,2\}\in\Gamma_{\min}$ and that every other member of
    $\Gamma_{\min}$ has cardinality $3$. Put
    $U=[n]\setminus\{1,2\}$, and define two graphs $G_1$ and $G_2$
    on the common vertex set $U$ by
    \[
    \begin{aligned}
    E(G_1)&:=\left\{\{u,v\}\in\binom{U}{2}:
        \{1,u,v\}\in\Gamma_{\min}\right\},\\
    E(G_2)&:=\left\{\{u,v\}\in\binom{U}{2}:
        \{2,u,v\}\in\Gamma_{\min}\right\}.
    \end{aligned}
    \]
	Then
	\[
	\Gamma_{\min}
	=
	\bigl\{\{1,2\}\bigr\}
	\cup
	\bigl\{\{1\}\cup e : e\in E(G_1)\bigr\}
	\cup
	\bigl\{\{2\}\cup f : f\in E(G_2)\bigr\}.
	\]
	Moreover, the edges of $G_2$ are precisely the
	inclusion-minimal vertex covers of $G_1$, and the edges of $G_1$
	are precisely the inclusion-minimal vertex covers of $G_2$.
\end{lemma}

\begin{proof}
	Let $M\in\Gamma_{\min}$ with $M\neq\{1,2\}$. Since any two
	authorized sets must intersect, $M$ contains at least one of
	the players $1$ and $2$. It cannot contain both, since it would then
	properly contain the minimal authorized set $\{1,2\}$. By
	assumption, $|M|=3$, so $M$ has one of the two forms appearing
	in the displayed decomposition.
	
    We next show that every edge of $G_2$ is an inclusion-minimal
    vertex cover of $G_1$. Let $f\in E(G_2)$. For every
    $e\in E(G_1)$, the two minimal authorized sets $\{2\}\cup f$
    and $\{1\}\cup e$ must intersect. Hence
    $e\cap f\neq\varnothing$, and therefore $f$ is a vertex cover
    of $G_1$.
	
	To prove inclusion-minimality, fix $x\in f$. Since
    $\{2\}\cup f$ is a minimal authorized set, its proper subset
    $\{2\}\cup\bigl(f\setminus\{x\}\bigr)$ is unauthorized. By
    complementarity, its complement in $[n]$, namely
    $\{1\}\cup\bigl(U\setminus(f\setminus\{x\})\bigr)$, is
    authorized and hence contains a minimal authorized set.
    Since this complement does not contain player $2$, the
    decomposition above implies that it contains a set
    $\{1\}\cup e_x$, where $e_x\in E(G_1)$. Consequently,
    $e_x\cap\bigl(f\setminus\{x\}\bigr)=\varnothing$. Thus
    $f\setminus\{x\}$ is not a vertex cover of $G_1$. This holds
    for every $x\in f$, so $f$ is inclusion-minimal.
	
	Conversely, let $C\subseteq U$ be an inclusion-minimal vertex
    cover of $G_1$. We first show that $\{2\}\cup C$ is authorized.
    Suppose, to the contrary, that it is unauthorized. By
    complementarity, $\{1\}\cup(U\setminus C)$ is authorized and
    hence contains a minimal authorized set.
    Since this set does not contain player $2$, it must contain a
    minimal authorized set of the form $\{1\}\cup e$, where
    $e\in E(G_1)$. It follows that $e\subseteq U\setminus C$,
    contradicting the assumption that $C$ meets every edge of
    $G_1$. Therefore $\{2\}\cup C$ is authorized.
	
	It remains to prove that $\{2\}\cup C$ is a minimal authorized set.
    First, $C$ is disjoint from the authorized set $\{1,2\}$, so
    $C$ is unauthorized. Next, fix $x\in C$. Since $C$ is an
    inclusion-minimal vertex cover, there exists an edge
    $e_x\in E(G_1)$ such that
    $e_x\cap\bigl(C\setminus\{x\}\bigr)=\varnothing$.
    The authorized set $\{1\}\cup e_x$ is therefore disjoint from
    $\{2\}\cup\bigl(C\setminus\{x\}\bigr)$, so the latter set is
    unauthorized. Thus every single-element deletion from
    $\{2\}\cup C$ is unauthorized. By monotonicity, every proper
    subset is unauthorized, and hence
    $\{2\}\cup C\in\Gamma_{\min}$.
	
	Since every member of $\Gamma_{\min}$ other than $\{1,2\}$ has
	cardinality $3$, it follows that $|C|=2$. Therefore
	$C\in E(G_2)$. We have proved that the edges of $G_2$ are
	exactly the inclusion-minimal vertex covers of $G_1$. The
	reverse statement follows by symmetry.
\end{proof}

\begin{theorem}\label{general}
	There is no pure perfect one-qubit-share QSS scheme with
	\(n\in\{3,4,6\}\) and no redundant players.
\end{theorem}

\begin{proof}
	\textbf{Cases $n=3,4$.}
	By Lemma~\ref{kbound}, any QSS scheme with $n = 3$ or $n = 4$ has minimal authorized sets of size at most $2$. No singleton can be a minimal authorized set under the no-redundant-players assumption: if \(\{i\}\) were authorized, then no minimal authorized set could contain any player other than \(i\), so all other players would be redundant. Hence every minimal authorized set has size \(2\), and the scheme is \(2\)-homogeneous. This case has already been ruled out by Theorem~\ref{mainthm}. Hence, no such scheme exists for $n = 3$ or $n = 4$.
	
	\textbf{Case $n=6$.}
	By Lemma~\ref{kbound}, every minimal authorized set has size at most \(3\). If all minimal authorized sets have the same cardinality, then the scheme is homogeneous and has already been ruled out in Theorem~\ref{mainthm}. Thus it remains to consider the non-homogeneous case.
	
	In the non-homogeneous case, there exists a minimal authorized set of size \(2\). By Lemma~\ref{notwo}, such a set is unique. Without loss of generality, assume that \(\{1,2\}\in\Gamma_{\min}\). Put \(U:=\{3,4,5,6\}\). Since every minimal authorized set must intersect \(\{1,2\}\), and no minimal authorized set other than \(\{1,2\}\) can contain both \(1\) and \(2\), every other minimal authorized set is of the form \(\{1\}\cup X\) or \(\{2\}\cup Y\), where \(X,Y\in\binom{U}{2}\). Define
	\[
	F_1:=\left\{X\in\binom{U}{2}:\{1\}\cup X\in\Gamma_{\min}\right\},
	\qquad
	F_2:=\left\{Y\in\binom{U}{2}:\{2\}\cup Y\in\Gamma_{\min}\right\}.
	\]
	We view \(F_1\) and \(F_2\) as graphs on the common vertex set \(U\). By Lemma~\ref{lem:minimal-pair-graphs}, the members of \(F_2\) are precisely the inclusion-minimal vertex covers of \(F_1\), and, symmetrically, the members of \(F_1\) are precisely the inclusion-minimal vertex covers of \(F_2\).
	
	We now classify the possible pairs \((F_1,F_2)\). Since the access structure has no redundant players, every vertex of \(U\) must appear in at least one edge of \(F_1\) or \(F_2\). Moreover, no vertex of \(U\) can be isolated in \(F_1\). Indeed, an isolated vertex cannot belong to an inclusion-minimal vertex cover of \(F_1\), since deleting it would not affect whether the remaining set meets every edge. Therefore, by Lemma~\ref{lem:minimal-pair-graphs}, such a vertex would also be isolated in \(F_2\). It would then occur in no minimal authorized set, contradicting non-redundancy. Thus \(F_1\) has no isolated vertices, and the same argument shows that \(F_2\) has no isolated vertices.
	
	Since the members of \(F_2\) are precisely the inclusion-minimal vertex covers of \(F_1\), and
	\[
	F_2\subseteq\binom{U}{2},
	\]
	every inclusion-minimal vertex cover of the graph \(F_1\) has size \(2\). A graph on four vertices with no isolated vertices and whose inclusion-minimal vertex covers all have size \(2\) is, up to isomorphism, one of \(2K_2\), \(P_4\), or \(C_4\). Indeed, if it has two edges, then the two edges must be disjoint, giving \(2K_2\). If it has three edges, the only possibilities without isolated vertices are \(P_4\) and \(K_{1,3}\), but \(K_{1,3}\) has an inclusion-minimal vertex cover of size \(1\). If it has four edges, the only admissible case is \(C_4\); the other four-edge graph contains a triangle with a pendant edge and has an inclusion-minimal vertex cover of size \(3\). Finally, graphs with five or six edges also have inclusion-minimal vertex covers of size \(3\). Therefore only \(2K_2\), \(P_4\), and \(C_4\) remain.
	
	The inclusion-minimal vertex covers of \(2K_2\), \(P_4\), and \(C_4\), viewed as edge families on the same vertex set, form graphs isomorphic to \(C_4\), \(P_4\), and \(2K_2\), respectively. Thus, up to relabeling the players in \(U\) and possibly swapping players \(1\) and \(2\), there are only two possible forms for \(\Gamma_{\min}\):
	\begin{equation}\label{ac1}
		\Gamma_{\min}
		=
		\{
		\{1,2\},
		\{1,3,5\},\{1,3,6\},\{1,4,5\},\{1,4,6\},
		\{2,3,4\},\{2,5,6\}
		\},
	\end{equation}
	or
	\begin{equation}\label{ac2}
		\Gamma_{\min}
		=
		\{
		\{1,2\},
		\{1,3,5\},\{1,4,5\},\{1,4,6\},
		\{2,3,4\},\{2,4,5\},\{2,5,6\}
		\}.
	\end{equation}
	
	We now rule out both cases using entropy inequalities. Let \(R\) denote the reference system, and write \(S(X)\) for the entropy of subsystem \(X\) in the associated QSS state. We shall use the following consequence of Proposition~\ref{QSS-state-MI}. If \(A\subseteq[6]\) is authorized, then \(S(A^c)=S(R\cup A)=S(A)-\log 2\), where \(A^c\) denotes the complement of \(A\) inside the player set \([6]\). If \(A\) is forbidden, then \(S(A^c)=S(R\cup A)=S(A)+\log 2\). Both identities use the purity of the global QSS state on \(R\cup[6]\).
	
	\medskip
	
	\noindent
	\textbf{Case I.}
	Assume that \(\Gamma_{\min}\) is given by Eq.~\eqref{ac1}. Apply strong subadditivity to the two subsystems \(\{1,3,4\}\) and \(\{3,4,5,6\}\):
	\[
	S(134)+S(3456)\ge S(13456)+S(34).
	\]
	We compute the four terms. Since \(\{2,5,6\}\in\Gamma_{\min}\), Corollary~\ref{min-log} gives \(S(256)=3\log 2\). As \(\{2,5,6\}\) is authorized, \(S(134)=S(R256)=S(256)-\log 2=2\log 2\). Similarly, since \(\{1,2\}\in\Gamma_{\min}\), we have \(S(12)=2\log 2\), and hence \(S(3456)=S(R12)=S(12)-\log 2=\log 2\). Moreover, the singleton \(\{2\}\) is forbidden and \(S(2)=\log 2\), so \(S(13456)=S(R2)=S(R)+S(2)=2\log 2\). Finally, since \(\{3,4\}\subseteq\{2,3,4\}\in\Gamma_{\min}\), Corollary~\ref{min-log} gives \(S(34)=2\log 2\). Substituting these values into strong subadditivity yields
	\[
	2\log 2+\log 2
	\ge
	2\log 2+2\log 2,
	\]
	which is impossible. Hence Case I cannot occur.
	
	\medskip
	
	\noindent
	\textbf{Case II.}
	Assume that \(\Gamma_{\min}\) is given by Eq.~\eqref{ac2}. Apply strong subadditivity to the two subsystems \(\{1,3,4\}\) and \(\{1,3,6\}\):
	\[
	S(134)+S(136)\ge S(1346)+S(13).
	\]
	Since \(\{2,5,6\}\in\Gamma_{\min}\), we have \(S(256)=3\log 2\), and therefore \(S(134)=S(R256)=S(256)-\log 2=2\log 2\). Similarly, since \(\{2,4,5\}\in\Gamma_{\min}\), \(S(136)=S(R245)=S(245)-\log 2=2\log 2\). Next, \(\{2,5\}\) is forbidden. Indeed, \(\{1,2\}\) is the unique minimal authorized set of size \(2\), so no other subset of size \(2\) can be authorized. Since \(\{2,5\}\subseteq\{2,5,6\}\in\Gamma_{\min}\), Corollary~\ref{min-log} gives \(S(25)=2\log 2\). Hence secrecy gives \(S(1346)=S(R25)=S(R)+S(25)=3\log 2\). Finally, since \(\{1,3\}\subseteq\{1,3,5\}\in\Gamma_{\min}\), Corollary~\ref{min-log} gives \(S(13)=2\log 2\). Substituting these values into strong subadditivity gives
	\[
	2\log 2+2\log 2
	\ge
	3\log 2+2\log 2,
	\]
	again impossible. Hence Case II cannot occur.
	
	Both possible non-homogeneous minimal access structures lead to contradictions. Therefore no pure perfect QSS scheme with no redundant players and a non-homogeneous access structure exists for \(n=6\). Combining this with the homogeneous classification proves the theorem.
\end{proof}
The following lemma is used to rule out the non-homogeneous five-player case.
\begin{lemma}
\label{18}
There is no $2$-uniform state of $6$ qubits with exactly $18$ three-qubit
subsystems whose reduced density matrices are maximally mixed.
\end{lemma}
\begin{proof}
We use Rains' shadow inequality, namely the \(k=2\) specialization of
Theorem~3 in~\cite{817508}: for any fixed subset \(T\subseteq[N]\),
\[
    s_T(\rho)=\sum_{S\subseteq [N]}(-1)^{|S\cap T|}
    \Tr\!\left(\rho_S^2\right)\geq 0.
\]
Suppose, for contradiction, that there exists a $6$-qubit $2$-uniform pure state
$\rho$ with exactly $18$ three-qubit subsystems whose reduced density matrices
are maximally mixed. There are $\binom{6}{3}=20$ three-qubit subsystems. By the
Schmidt decomposition, complementary reductions of a pure state have the same
nonzero eigenvalues and hence the same purity. Without loss of generality, let
$\{1,2,3\}$ and $\{4,5,6\}$ be the two three-qubit subsystems whose reductions
are not maximally mixed. Then
\[
    \rho_A=\frac{\mathbb{I}_{2^{|A|}}}{2^{|A|}}
    \quad\text{for every } A\in\binom{[6]}{2},
\]
and
\[
    \rho_A=\frac{\mathbb{I}_{2^3}}{2^3}
    \quad\text{for every }
    A\in\binom{[6]}{3}\setminus\{\{1,2,3\},\{4,5,6\}\}.
\]
Let $x=\Tr\!\left(\rho_S^2\right)$ for
$S\in\{\{1,2,3\},\{4,5,6\}\}$, and choose $T=\{1,2,3,4\}$.
We now evaluate \(s_T(\rho)\). For subsets of size \(m\), the signed count is
\[
    c_m=\sum_{|S|=m}(-1)^{|S\cap T|}
    =[z^m](1-z)^4(1+z)^2.
\]
Thus \((c_0,\ldots,c_6)=(1,-2,-1,4,-1,-2,1)\). The two exceptional triples
\(\{1,2,3\}\) and \(\{4,5,6\}\) both have sign \(-1\), so the non-exceptional
triples have signed count \(4+2=6\). The contributions to \(s_T(\rho)\) are
\[
\begin{array}{c|c|c|c}
\text{subsets }S & \text{signed count} & \Tr(\rho_S^2)
& \text{contribution}\\
\hline
|S|=0 & 1 & 1 & 1\\
|S|=1 & -2 & 1/2 & -1\\
|S|=2 & -1 & 1/4 & -1/4\\
|S|=3,\ S\text{ exceptional} & -2 & x & -2x\\
|S|=3,\ S\text{ non-exceptional} & 6 & 1/8 & 3/4\\
|S|=4 & -1 & 1/4 & -1/4\\
|S|=5 & -2 & 1/2 & -1\\
|S|=6 & 1 & 1 & 1
\end{array}
\]
Here the purities for \(m=4,5,6\) follow from purity of the global state and the
corresponding complementary reductions of sizes \(2,1,0\). Summing the last
column gives
\[
    s_T(\rho)=-2x+\frac14.
\]
Since $\rho_S$ is not maximally mixed for
$S\in\{\{1,2,3\},\{4,5,6\}\}$, its eigenvalues $\lambda_1,\ldots,\lambda_8$ are
not all equal to $1/8$. Therefore
\[
    x=\Tr\!\left(\rho_S^2\right)=\sum_{i=1}^{8}\lambda_i^2>\frac18.
\]
This implies $s_T(\rho)<0$, contradicting the shadow inequality. Hence such a
state does not exist.
\end{proof}

\begin{theorem}\label{5}
	There is no pure perfect one-qubit-share QSS scheme with no redundant
players and a non-homogeneous access structure when the number of players is
\(n=5\).
\end{theorem}

\begin{proof}
Suppose that such a non-homogeneous scheme exists. By Lemma~\ref{kbound}, every
minimal authorized set has size \(2\) or \(3\), and non-homogeneity implies that
both sizes occur. By Lemma~\ref{notwo}, there is at most one minimal authorized
set of size \(2\). Relabel the players so that this unique pair is
\(\{1,2\}\in\Gamma_{\min}\).

There is also a minimal authorized set of size \(3\). It must intersect
\(\{1,2\}\), and it cannot contain both \(1\) and \(2\), since otherwise it would
properly contain the minimal authorized set \(\{1,2\}\). After possibly
interchanging \(1\) and \(2\) and relabeling the remaining players, we may
assume that \(\{1,3,4\}\in\Gamma_{\min}\).

We now determine the entire minimal access structure. Every subset of size \(2\) other
than \(\{1,2\}\) is unauthorized, because it is not the unique minimal
authorized pair and it cannot contain a minimal authorized triple. Hence the
complement of each such pair is authorized. If a pair \(P\neq\{1,2\}\) meets
\(\{1,2\}\) in exactly one element, then \(P^c\) is a subset of size \(3\) not
containing \(\{1,2\}\). Since \(P^c\) is authorized and cannot contain the
unique minimal authorized pair, it must itself be minimal authorized. This gives
the six triples
\[
\{1,3,4\},\{1,3,5\},\{1,4,5\},
\{2,3,4\},\{2,3,5\},\{2,4,5\}.
\]
Conversely, a subset of size \(3\) containing \(\{1,2\}\) is not minimal, and
\(\{3,4,5\}\) is unauthorized because its complement \(\{1,2\}\) is authorized.
Thus
\[
\Gamma_{\min}
=
\bigl\{
\{1,2\}, \{1,3,4\}, \{1,3,5\}, \{1,4,5\},
\{2,3,4\}, \{2,3,5\}, \{2,4,5\}
\bigr\}.
\]

Let \(\ket{\psi_{\mathcal S}}\) be the associated six-qubit QSS state, with
reference system \(R\), and let \(\rho=\ketbra{\psi_{\mathcal S}}{\psi_{\mathcal S}}\).
We first show that \(\rho\) is \(2\)-uniform. If a two-qubit subsystem is
\(R\cup\{i\}\), then \(\{i\}\) is unauthorized, so secrecy gives
\[
    S(Ri)=S(R)+S(i)=2\log 2,
\]
where \(S(i)=\log 2\) follows from Corollary~\ref{coro1}. If a two-qubit
subsystem is a player pair, then the displayed form of \(\Gamma_{\min}\) shows
that it is contained in some minimal authorized set. Corollary~\ref{min-log}
therefore gives entropy \(2\log 2\) for every player pair. Hence every two-qubit
reduction is maximally mixed.

It remains to count the maximally mixed three-qubit reductions. Among the ten
triples containing \(R\), only \(R\cup\{1,2\}\) contains an authorized player
pair. For this triple, recoverability gives
\[
    S(R12)=S(12)-S(R)=\log 2,
\]
so it is not maximally mixed. For the other nine triples \(R\cup P\), the pair
\(P\) is unauthorized, and secrecy gives
\[
    S(RP)=S(R)+S(P)=3\log 2.
\]
Thus exactly nine triples containing \(R\) are maximally mixed.

Now consider the ten player triples. If a player triple \(T\) is authorized,
then its complement \(T^c\) is an unauthorized player pair. Since the global
state is pure, and since secrecy gives \(S(R\cup T^c)=S(R)+S(T^c)\),
\[
    S(T)=S(R\cup T^c)=S(R)+S(T^c)=3\log 2.
\]
The only unauthorized player triple is \(\{3,4,5\}\), because its complement is
the unique authorized player pair \(\{1,2\}\). For this triple,
\[
    S(345)=S(R12)=\log 2,
\]
so it is not maximally mixed. Hence exactly nine player triples are maximally
mixed.

Therefore \(\rho\) is a \(2\)-uniform six-qubit state with exactly
\(9+9=18\) maximally mixed three-qubit reductions. This contradicts
Lemma~\ref{18}. Thus no pure perfect QSS scheme with no redundant players and a
non-homogeneous access structure exists for \(n=5\).
\end{proof}
We next prove that there is no non-homogeneous QSS scheme with $7$ players.
We first rule out the existence of a minimal authorized set of size $4$.
\begin{theorem}\label{thm:no-nonhomogeneous-n7}
There is no pure perfect one-qubit-share QSS scheme with no redundant players
and a non-homogeneous access structure when the number of players is \(7\).
\end{theorem}

\begin{proof}
Assume for contradiction that such a scheme exists.

We first prove that there is no minimal authorized set of size \(4\). Suppose otherwise.
Without loss of generality, assume that
\(A=\{1,2,3,4\}\in\Gamma_{\min}\), and let \(|\psi_{\mathcal{S}}\rangle\) be the associated
QSS state. Since \(A\) is a minimal authorized set, every proper subset of
\(A\) is unauthorized.
By Corollary~\ref{min-log}, we have \(S(B)=|B|\log 2\) for every \(B\subseteq A\). In
particular, \(S(1234)=4\log 2\), and \(S(C)=3\log 2\) for every
\(C\in\binom{A}{3}\).

Now fix any \(C\in\binom{A}{3}\). Since \(C\) is unauthorized, the secrecy
condition~\eqref{sec} gives \(I(R:C)=0\). Hence
\(S(R\cup C)=S(R)+S(C)=4\log 2\).
Therefore the five four-qubit subsystems
$\{1234, R123, R124, R134,R234\}$
are all maximally mixed. Let \(W=\{R,1,2,3,4\}\). Then every subset of size \(4\)
contained in \(W\) is maximally mixed, contradicting Lemma~\ref{Zhang}, which
states that for any five-qubit subset of an eight-qubit pure
state, at least one of its subsets of size \(4\) is not maximally mixed.
Thus no minimal authorized set of size \(4\) can exist.

By Lemma~\ref{kbound}, every minimal authorized set has size at most \(4\). Since minimal
authorized sets of size \(4\) have just been ruled out, every minimal authorized set has size \(2\) or \(3\). The
access structure is assumed to be non-homogeneous, so there must exist a minimal authorized
set of size \(2\). By Lemma~\ref{notwo}, such a set is unique. Without loss of generality,
assume that \(\{1,2\}\in\Gamma_{\min}\). Hence every other minimal authorized set has size
\(3\).

Put \(U:=\{3,4,5,6,7\}\). Every minimal authorized set other than \(\{1,2\}\) must
intersect \(\{1,2\}\), and cannot contain both \(1\) and \(2\), since \(\{1,2\}\) is already
a minimal authorized set. Therefore every such set is of the form \(\{1\}\cup X\) or
\(\{2\}\cup Y\), where \(X,Y\in\binom{U}{2}\). Define
\[
    \mathcal F_1
    :=
    \{X\in\binom{U}{2}: \{1\}\cup X\in\Gamma_{\min}\},
    \qquad
    \mathcal F_2
    :=
    \{Y\in\binom{U}{2}: \{2\}\cup Y\in\Gamma_{\min}\}.
\]
Then
\[
    \Gamma_{\min}
    =
    \{\{1,2\}\}
    \cup
    \{\{1\}\cup X:X\in\mathcal F_1\}
    \cup
    \{\{2\}\cup Y:Y\in\mathcal F_2\}.
\]
We now view \(\mathcal F_1\) and \(\mathcal F_2\) as graphs on the common vertex set
\(U\). By Lemma~\ref{lem:minimal-pair-graphs}, the members of \(\mathcal F_2\) are
precisely the inclusion-minimal vertex covers of \(\mathcal F_1\), and, symmetrically,
the members of \(\mathcal F_1\) are precisely the inclusion-minimal vertex covers of
\(\mathcal F_2\).

We claim that neither graph has an isolated vertex. Indeed, suppose that \(v\in U\) is
isolated in \(\mathcal F_1\). Such a vertex cannot belong to an inclusion-minimal vertex
cover of \(\mathcal F_1\), since deleting \(v\) would not affect whether the remaining set
meets every edge. Therefore, by Lemma~\ref{lem:minimal-pair-graphs}, \(v\) is also isolated
in \(\mathcal F_2\). It then appears in neither graph, and consequently the player \(v\)
appears in no minimal authorized set, contradicting the non-redundancy condition. Thus
\(\mathcal F_1\) has no isolated vertices. The same argument shows that \(\mathcal F_2\)
has no isolated vertices.

Finally, every inclusion-minimal vertex cover of \(\mathcal F_1\) is a member of
\(\mathcal F_2\subseteq\binom{U}{2}\), and hence has size \(2\). We show that this is
impossible for a graph on five vertices with no isolated vertices. Indeed, suppose that a
graph \(G\) on five vertices has no isolated vertices and that all its inclusion-minimal
vertex covers have size \(2\).
Equivalently, all maximal independent sets of \(G\) have size \(3\). Then every vertex must
have at least two non-neighbors, since any vertex can be extended to a maximal independent
set of size \(3\). Hence every vertex has degree at most \(2\). Since \(G\) has no isolated
vertices, all degrees are \(1\) or \(2\), so \(G\) is a disjoint union of paths and cycles.

There are only four possible forms on five vertices with no isolated vertices:
\(C_5\), \(C_3\cup P_2\), \(P_5\), and \(P_3\cup P_2\). Each of them has a maximal
independent set of size \(2\): for \(C_5\), take two vertices at distance \(2\); for
\(C_3\cup P_2\), take one vertex from the triangle and one endpoint of the edge; for
\(P_5\), take the second and fourth vertices along the path; and for \(P_3\cup P_2\), take
the middle vertex of \(P_3\) and one endpoint of \(P_2\). This contradicts the assumption
that all maximal independent sets have size \(3\).

Applying this elementary graph fact to \(\mathcal F_1\), we obtain a contradiction. Therefore
no such non-homogeneous pure perfect QSS scheme with seven players and no
redundant players exists.
\end{proof}

\begin{proof}[Proof of Theorem~\ref{thm:classification-n-leq-7}]
The homogeneous cases are classified by Theorem~\ref{mainthm}.
Theorem~\ref{general} excludes the cases \(n=3,4,6\), while
Theorems~\ref{5} and~\ref{thm:no-nonhomogeneous-n7} exclude additional
non-homogeneous access structures for \(n=5\) and \(n=7\), respectively.
Combining these results proves the stated classification.
\end{proof}

	\section{Concluding Remarks}\label{c}

This work addresses two related questions about ideal qubit quantum secret
sharing. The first asks how far the AME--threshold correspondence extends
beyond threshold access structures. We prove that every pure perfect
\(3\)-homogeneous one-qubit-share QSS scheme with no redundant players gives a
\(3\)-uniform QSS state, but the two counterexamples in
Examples~\ref{ex:uniform-not-qss} and~\ref{ex:k-uniform-counterexample} show
that there is no general equivalence between \(k\)-uniform states and
non-threshold QSS states. Under our convention that ``QSS state'' refers only
to the Choi state of a pure perfect QSS scheme, a \(k\)-uniform state need not
be a QSS state, and a \(k\)-homogeneous non-threshold ideal qubit QSS scheme
need not induce a \(k\)-uniform QSS state.

The second question concerns resource-optimal realizability of access
structures. Although general QSS constructions can realize broad classes of
access structures satisfying the usual quantum constraints, they may require
large shares. In the one-qubit-share regime, the access structure becomes much
more rigid. We capture part of this rigidity through restrictions derived from
complementarity, non-redundancy, and entropy inequalities, including bounds on
minimal authorized sets and constraints on authorized pairs. These tools lead
to a complete classification for at most seven players: the only realizable
minimal access structures are the \(3\)-threshold structure for \(n=5\) and the
Fano-plane structure for \(n=7\).

For larger numbers of players, the main difficulty is that the class of
combinatorially admissible minimal access structures grows quickly, while
entropy constraints become harder to exploit in a purely combinatorial way. A
natural direction is to combine direct access-structure analysis and matroidal
viewpoints with stabilizer-code constructions and stronger entropy inequalities.
Such a framework may lead to broader classification results and to a sharper
understanding of when uniform-state methods genuinely capture non-threshold
ideal QSS schemes.

\section*{Acknowledgment}
This work was supported by the Quantum Science and Technology--National Science and Technology Major Project (QNMP), Project No. 2021ZD0302901 (Topic 1). We also acknowledge support from Hefei National Laboratory. We thank Dr. Ningyu for many helpful discussions throughout this project.

\section*{Conflict of Interest}
The authors declare no conflicts of interest.

\appendices
\section{Explicit Forms of the QSS Examples}
\label{app:explicit-examples}

In this appendix, we give the explicit sharing maps for the QSS schemes in
Examples~\ref{thres} and~\ref{homog}, together with the associated QSS states.

\subsection{The $3$-threshold QSS scheme with five players}

The sharing map in Example~\ref{thres} is the linear map
$\Lambda_s:\mathbb{C}^2\rightarrow(\mathbb{C}^2)^{\otimes 5}$ defined by
\[
\begin{aligned}
|0\rangle \mapsto \frac{1}{4}(&
|00000\rangle+|10010\rangle+|01001\rangle+|10100\rangle
+|01010\rangle \\
&-|11011\rangle-|00110\rangle-|11000\rangle
-|11101\rangle-|00011\rangle\\
&-|11110\rangle-|01111\rangle
-|10001\rangle-|01100\rangle-|10111\rangle+|00101\rangle),
\end{aligned}
\]
and
\[
\begin{aligned}
|1\rangle \mapsto \frac{1}{4}(&
|11111\rangle+|01101\rangle+|10110\rangle+|01011\rangle
+|10101\rangle\\
&-|00100\rangle-|11001\rangle-|00111\rangle
-|00010\rangle-|11100\rangle\\
&-|00001\rangle-|10000\rangle
-|01110\rangle-|10011\rangle-|01000\rangle+|11010\rangle).
\end{aligned}
\]
This encoding map is the $5$-qubit quantum error-correcting code
\cite{gottesman2009introductionquantumerrorcorrection}, and it gives a pure
$3$-threshold QSS scheme with five players~\cite{PhysRevLett.83.648}.

The QSS state associated with the scheme in Example~\ref{thres} is
\[
\begin{aligned}
|\psi_{\mathrm{th}}\rangle
=
\frac{1}{4}(&
\ket{000}(\ket{+-+}+\ket{-+-})
+\ket{001}(-\ket{+--}+\ket{-++})\\
&+\ket{010}(\ket{++-}-\ket{--+})
+\ket{011}(-\ket{+++}-\ket{---})\\
&+\ket{100}(-\ket{+++}+\ket{---})
+\ket{101}(-\ket{++-}-\ket{--+})\\
&+\ket{110}(-\ket{+--}-\ket{-++})
+\ket{111}(-\ket{+-+}+\ket{-+-})
),
\end{aligned}
\]
where
\[
    \ket{+}=\frac{\ket{0}+\ket{1}}{\sqrt{2}},
    \qquad
    \ket{-}=\frac{\ket{0}-\ket{1}}{\sqrt{2}}.
\]

\subsection{The $3$-homogeneous QSS scheme from the Steane code}

The sharing map in Example~\ref{homog} is the linear map
$\Lambda_s:\mathbb{C}^2\rightarrow(\mathbb{C}^2)^{\otimes 7}$ defined by
\[
\begin{aligned}
|0\rangle \mapsto \frac{1}{2\sqrt{2}}(&
|0000000\rangle+|1000111\rangle+|0101011\rangle+|0011110\rangle\\
&+|1101100\rangle+|1011001\rangle+|0110101\rangle+|1110010\rangle),
\end{aligned}
\]
and
\[
\begin{aligned}
|1\rangle \mapsto \frac{1}{2\sqrt{2}}(&
|1111111\rangle+|0111000\rangle+|1010100\rangle+|1100001\rangle\\
&+|0010011\rangle+|0100110\rangle+|1001010\rangle+|0001101\rangle).
\end{aligned}
\]
This encoding map is the Steane code~\cite{steane1996multiple}, and it gives
the pure $3$-homogeneous QSS scheme with seven players described in
Example~\ref{homog}~\cite{PhysRevA.81.052333}.

The QSS state associated with the scheme in Example~\ref{homog} is
\[
\begin{aligned}
|\psi_{\mathrm{St}}\rangle
=
\frac{1}{4}(&
\ket{00000000}+\ket{00001111}
+\ket{00110011}+\ket{00111100}\\
&+\ket{01010101}+\ket{01011010}
+\ket{01100110}+\ket{01101001}\\
&+\ket{10010110}+\ket{10011001}
+\ket{10100101}+\ket{10101010}\\
&+\ket{11000011}+\ket{11001100}
+\ket{11110000}+\ket{11111111}).
\end{aligned}
\]
\section{Proofs of Known Results}
\label{app:known-results}

In this appendix, we provide proofs of the known entropic results collected in
Section~\ref{sec:pre}.
\subsection{Proof of Theorem~\ref{single}}

\begin{proof}
Since $A\cup B\in\Gamma$, Proposition~\ref{QSS-state-MI} gives
\[
    S(AB)-S(RAB)=S(R).
\]
Applying the Araki--Lieb inequality to $S(RAB)$ yields
\[
    S(AB)-S(RA)+S(B)\geq S(R).
\]
Since $I(A:R)=0$, it follows that $S(RA)=S(A)+S(R)$, and this together with
the last inequality gives
\[
    S(AB)-S(A)+S(B)\geq 2S(R).
\]
Using subadditivity and the fact that $S(R)=\log 2$, we obtain
\[
    S(B)\geq \log 2.
\]
By symmetry, we also have $S(A)\geq \log 2$. This proves the theorem.
\end{proof}

\subsection{Proof of Corollary~\ref{coro1}}

\begin{proof}
Since \(i\) is non-redundant, there exists
\(B\subseteq[n]\setminus\{i\}\) such that \(B\notin\Gamma\) and
\(B\cup\{i\}\in\Gamma\).

If \(\{i\}\) is authorized, then recoverability gives
\(I(R:\{i\})=2\log 2\). Since \(S(R)=\log 2\), \(S(i)\leq\log 2\), and
entropy is nonnegative, the identity
\[
    I(R:\{i\})=S(R)+S(i)-S(Ri)
\]
forces \(S(i)=\log 2\).

It remains to consider the case in which \(\{i\}\) is unauthorized. Applying
Theorem~\ref{single} to the unauthorized sets \(\{i\}\) and \(B\), whose union
\(B\cup\{i\}\) is authorized, gives
\[
    S(i)\geq \log 2.
\]
On the other hand, since each share is one qubit,
\[
    S(i)\leq \log 2.
\]
Therefore, $S(i)=\log 2$.
\end{proof}

\subsection{Proof of Lemma~\ref{entropy}}

\begin{proof}
Let $R$ be the reference system of the associated QSS state. Since $B$ is
forbidden, the secrecy condition gives $I(R:B)=0$, and hence
\[
    S(RB)=S(R)+S(B).
\]
Since $B\cup\{i\}$ is authorized, the recoverability condition gives
$I(R:Bi)=2S(R)$, or equivalently,
\[
    S(RBi)=S(Bi)-S(R).
\]
By the Araki--Lieb inequality, $S(RBi)\geq S(RB)-S(i)$. Combining the last
three relations, we obtain
\[
    S(Bi)-S(R)
    =
    S(RBi)
    \geq
    S(RB)-S(i)
    =
    S(B)+S(R)-S(i).
\]
The conditions \(B\notin\Gamma\) and \(B\cup\{i\}\in\Gamma\) imply that
player \(i\) is non-redundant. By Corollary~\ref{coro1},
\(S(i)=S(R)=\log 2\). Therefore
$S(Bi)\geq S(B)+S(i)$. The reverse inequality follows from subadditivity, so
\[
    S(Bi)=S(B)+S(i).
\]

Now let $X\subseteq B$. By strong subadditivity applied to the pair of
subsystems $iX$ and $B$, we have
\[
    S(iX)+S(B)\geq S(iB)+S(X),
\]
and hence
\[
    S(iX)-S(X)\geq S(iB)-S(B)=S(i).
\]
On the other hand, subadditivity gives $S(iX)\leq S(i)+S(X)$. Thus
$S(iX)=S(i)+S(X)$, i.e.,
\[
    S(\{i\}\cup X)=S(i)+S(X).
\]
\end{proof}

\subsection{Proof of Corollary~\ref{min-log}}

\begin{proof}
If $|B|=1$, its unique element belongs to the minimal authorized set \(A\), and
hence is non-redundant; the claim follows from Corollary~\ref{coro1}. Now assume
$|B|\geq 2$ and choose $i\in B$. Since $A$ is a minimal authorized set,
$A\setminus\{i\}$ is unauthorized. Applying Lemma~\ref{entropy} with
$X=B\setminus\{i\}$ gives
\[
    S(B)
    =
    S(i)+S(B\setminus\{i\})
    =
    \log 2+S(B\setminus\{i\}).
\]
Iterating this identity yields $S(B)=|B|\log 2$.
\end{proof}

\subsection{Proof of Lemma~\ref{Zhang}}

\begin{proof}
We prove the statement by contradiction. Let $n=4m$ with $m\geq 2$, and
suppose that there exists a subset $A\subseteq[n]$ with $|A|=2m+1$ such that
the reduction of $|\psi\rangle$ to every subset of size $2m$ contained in $A$ is maximally mixed.
Without loss of generality, assume that $A=[2m+1]$.

We first recall the notation used in the Bloch representation. For a state
$\rho$ on $t$ qubits, every operator can be expanded in the Pauli basis. We
write
\[
    \rho
    =
    \frac{1}{2^t}
    \left(
        \mathbb{I}+\sum_{j=1}^t P_j
    \right),
\]
where $P_j$ denotes the sum of all Pauli terms of weight $j$, i.e., tensor
products of single-qubit Pauli matrices with exactly $j$ non-identity factors.
We shall also use the following parity rule for Pauli strings: if two Pauli
strings of weights $p$ and $q$ have a nonzero anticommutator, then every Pauli
string appearing in that anticommutator has weight congruent to $p+q$ modulo
$2$.

Since every subset of size $2m$ contained in $A$ has maximally mixed reduction, all Pauli
coefficients in $\rho_A$ supported on at most $2m$ qubits must vanish. Indeed,
if a nonzero Pauli term had support $J\subseteq A$ with $|J|\leq 2m$, then it
would survive in the reduction to some subset of size $2m$ contained in $A$ and containing $J$,
contradicting maximal mixedness of that reduction. Hence the only possible
nonzero non-identity terms in the Bloch expansion of $\rho_A$ have full weight
$2m+1$, and therefore
\[
    \rho_A
    =
    \frac{1}{2^{2m+1}}
    \left(
        \mathbb{I}_A+P_{2m+1}
    \right),
\]
where $P_{2m+1}$ denotes the sum of all full-weight Pauli terms supported on
$A$.

Next, we derive a spectral constraint on $\rho_A$. Since the reduction to
$[2m]\subseteq A$ is maximally mixed and the global state is pure, the reduction
to its complement $[4m]\setminus[2m]$ is also maximally mixed. By tracing out
the subsystem $2m+1$, it follows that the reduction to
$A^c=[4m]\setminus A$ is maximally mixed. Since $|A^c|=2m-1$, the reduced state
$\rho_{A^c}$ has all $2^{2m-1}$ eigenvalues equal to $2^{1-2m}$. By the
Schmidt decomposition across the bipartition $A|A^c$, the nonzero eigenvalues
of $\rho_A$ and $\rho_{A^c}$ coincide. Therefore $\rho_A$ is proportional to a
projector and satisfies
\[
    \rho_A^2=2^{1-2m}\rho_A.
\]

Substituting the Bloch expansion of $\rho_A$ into this identity gives
\[
    (\mathbb{I}_A+P_{2m+1})^2=4(\mathbb{I}_A+P_{2m+1}).
\]
Now compare the terms of odd Pauli weight on both sides. The term
$2P_{2m+1}$ on the left-hand side has odd weight $2m+1$. On the other hand,
every term arising from $P_{2m+1}^2$ is obtained from products of Pauli strings
of weight $2m+1$; by the parity rule, such terms have even weight whenever
they contribute through a nonzero anticommutator. Hence $P_{2m+1}^2$ contributes
no odd-weight terms. Therefore the odd-weight part of the above equation gives
\[
    2P_{2m+1}=4P_{2m+1}.
\]
Thus $P_{2m+1}=0$, and consequently
\[
    \rho_A=\frac{\mathbb{I}_A}{2^{2m+1}}.
\]
This is impossible, because we already showed from the Schmidt decomposition
that $\rho_A$ has rank at most $2^{2m-1}$, whereas $\mathbb{I}_A/2^{2m+1}$ has full rank
$2^{2m+1}$. This contradiction proves that for every
$A\subseteq[n]$ with $|A|=2m+1$, at least one subset $B\subseteq A$ of size
$2m$ has non-maximally mixed reduction.
\end{proof}

\subsection{Proof of Lemma~\ref{two}}

\begin{proof}
Let $R$ denote the reference system in the associated QSS state
$|\psi_{\mathcal{S}}\rangle$. Since the secret system is a single qubit, we have
$S(R)=\log 2$.

We first record a simple consequence of the mutual-information characterization
of QSS states. If a subset $X\subseteq[n]$ is authorized, then
$I(R:X)=2\log 2$, and hence
\[
    S(R\cup X)
    =
    S(R)+S(X)-I(R:X)
    =
    S(X)-\log 2.
\]
If $X$ is forbidden, then $I(R:X)=0$, and hence
\[
    S(R\cup X)
    =
    S(R)+S(X)
    =
    S(X)+\log 2.
\]

Now let $A,B\in\Gamma$ be authorized sets such that $A\cap B\notin\Gamma$.
Since the access structure is monotone, $A\cup B$ is also authorized. Applying
strong subadditivity to the two subsystems $R\cup A$ and $R\cup B$ gives
\[
    S(R\cup A)+S(R\cup B)
    \geq
    S(R\cup A\cup B)+S(R\cup A\cap B).
\]
Because $A$ and $B$ are authorized, the identities above give
$S(R\cup A)=S(A)-\log 2$ and $S(R\cup B)=S(B)-\log 2$. Since $A\cup B$ is
authorized, $S(R\cup A\cup B)=S(A\cup B)-\log 2$. Since $A\cap B$ is forbidden,
$S(R\cup A\cap B)=S(A\cap B)+\log 2$. Substituting these four identities into
the strong subadditivity inequality yields
\[
    S(A)-\log 2+S(B)-\log 2
    \geq
    S(A\cup B)-\log 2+S(A\cap B)+\log 2.
\]
Rearranging gives
\[
    S(A)+S(B)
    \geq
    S(A\cup B)+S(A\cap B)+2\log 2.
\]
This proves the lemma.
\end{proof}

\section{Verification of the Self-Dual-Code Example}
\label{app:self-dual-code-verification}

We record the finite verification used in Example~\ref{ex:k-uniform-counterexample}.
The accompanying script
\texttt{verify\_self\_dual\_code\_example.py}, provided as supplementary
material, reproduces all computations below using only elementary linear
algebra over \(\mathbb F_2\).

Let \(g_1,\ldots,g_9\) be the rows of the matrix \(G\) in
Example~\ref{ex:k-uniform-counterexample}. Since \(G=[\mathbb{I}_9\mid A]\), it is enough
to check \(AA^T=\mathbb{I}_9\) over \(\mathbb F_2\). The script verifies this
equivalently by checking \(g_i\cdot g_j=0\) for all \(1\leq i,j\leq 9\), and it
also verifies \(\rank(G)=9\). Hence the binary length-\(18\) code
\(C=\langle g_1,\ldots,g_9\rangle\) is self-dual.

The weight enumerator is obtained by enumerating all row sums
\[
    c(\epsilon)=\sum_{i=1}^9 \epsilon_i g_i,
    \qquad
    \epsilon=(\epsilon_1,\ldots,\epsilon_9)\in\mathbb F_2^9,
\]
and counting \(\wt(c(\epsilon))\). This gives
\[
    W_C(y)=1+17y^4+51y^6+187y^8+187y^{10}
    +51y^{12}+17y^{14}+y^{18}.
\]

For the dealer/reference coordinate \(R=14\), the same enumeration restricted
to codewords whose support contains \(R\) gives
\[
    51y^6+187y^{10}+17y^{14}+y^{18}.
\]
The dealer-relative minimal supports are computed as follows. Let
\[
    \mathcal S_R=\{\supp(c):c\in C,\ R\in\supp(c)\}.
\]
A support \(S\in\mathcal S_R\) is retained precisely when there is no
\(T\in\mathcal S_R\) with \(T\subsetneq S\). Applying this inclusion test gives
exactly \(51\) retained supports, all of weight \(6\). Removing \(R\) from these
supports therefore gives \(51\) minimal authorized sets, all of size \(5\).

The no-redundant-player claim is certified by the five dealer-relative minimal
supports displayed in Example~\ref{ex:k-uniform-counterexample}. The script
checks that each of them is one of the \(51\) minimal supports and that their
union, after removing \(R=14\), is the full player set.

Finally, the non-\(5\)-uniform witness is explicit: the first generator row is
\[
    100000000000100011,
\]
whose support is \(B=\{1,13,17,18\}\). Thus \(C_B\) contains a nonzero codeword;
the script verifies in fact that \(C_B=\{0,g_1\}\), so \(\dim C_B=1\).
\begingroup
\renewcommand{\baselinestretch}{0.96}
\bibliographystyle{IEEEtran}
\bibliography{eo}
\endgroup

\end{document}